\documentclass[aps,physrev,amsmath,amssymb,reprint, superscriptaddress]{revtex4-2}
\usepackage{graphicx}
\usepackage{float}
\usepackage{hyperref}
\usepackage[capitalise]{cleveref}
\usepackage{bbm}
\usepackage{tikz}
\usetikzlibrary{quantikz2}

\begin{document}

\title{Unlocking early fault-tolerant quantum computing with mitigated magic dilution}

\date{August 8, 2025}

\author{Surabhi Luthra}
\email{zcaplut@ucl.ac.uk}
\affiliation{Department of Physics and Astronomy, University College London, London, WC1E 6BT, United Kingdom}

\author{Alexandra E. Moylett}
\altaffiliation[Current address: ]{Nu Quantum, 21 JJ Thompson Avenue, Cambridge, CB3 0FA, United Kingdom}
\affiliation{Riverlane, St Andrews House, 59 St Andrews Street, Cambridge CB2 3BZ, United Kingdom}

\author{Dan E. Browne}
\affiliation{Department of Physics and Astronomy, University College London, London, WC1E 6BT, United Kingdom}

\author{Earl T. Campbell}
\affiliation{Riverlane, St Andrews House, 59 St Andrews Street, Cambridge CB2 3BZ, United Kingdom}
\affiliation{Department of Physics and Astronomy, University of Sheffield, Sheffield, S3 7RH, United Kingdom}

\begin{abstract}

As quantum computing progresses towards the early fault-tolerant regime, quantum error correction will play a crucial role in protecting qubits and enabling logical Clifford operations. However, the number of logical qubits will initially remain limited, posing challenges for resource-intensive tasks like magic state distillation. It is therefore essential to develop efficient methods for implementing non-Clifford operations, such as small-angle rotations, to maximise the computational capabilities of devices within these constraints. In this work, we introduce mitigated magic dilution (MMD) as an approach to synthesise small-angle rotations by employing quantum error mitigation techniques to sample logical Clifford circuits given noisy encoded magic states. We explore the utility of our approach for the simulation of the 2D Fermi-Hubbard model. We identify evolution time regimes where MMD outperforms state-of-the-art synthesis techniques in the number of noisy encoded magic states required for square lattices up to size $8 \times 8$. Moreover, we demonstrate that our method can provide a practical advantage that is quantified by a better-than-quadratic improvement in the resource requirements for small-angle rotations over classical simulators. This work paves the way for early fault-tolerant demonstrations on devices supporting millions of quantum operations, the so-called MegaQuOp regime.

\end{abstract}

\maketitle

\section{\label{sec:introduction} Introduction}

Recent progress in experimental demonstrations of quantum error correction and logical computation \cite{acharyaQuantumErrorCorrection2024, bluvsteinLogicalQuantumProcessor2024, rodriguezExperimentalDemonstrationLogical2024, reichardtDemonstrationQuantumComputation2024, campbell2024series} has encouraged research towards practical applications of early fault-tolerant quantum computers \cite{campbellEarlyFaulttolerantSimulations2022, katabarwaEarlyFaultTolerantQuantum2024,toshioPracticalQuantumAdvantage2024, preskillBeyondNISQMegaquop2025}. The Gottesman-Knill theorem \cite{gottesmanHeisenbergRepresentationQuantum1998} shows that universal quantum computation necessitates non-Clifford gates. Implementing these gates typically requires the preparation of magic states or non-stabiliser states where resource-intensive techniques such as magic state distillation \cite{bravyiUniversalQuantumComputation2005} are used to improve their fidelity. Despite ongoing research in reducing the associated overhead of these factories \cite{litinskiMagicStateDistillation2019, gidney2019efficient, MSD24}, resource analysis of fault-tolerant implementations of large algorithms using magic state distillation require millions of physical qubits \cite{ogormanQuantumComputationRealistic2017,TensorHypercontraction, gidney2025factor, georges2025quantum}. In the anticipated early fault-tolerant era, where useful magic state distillation factories cannot be accommodated due to physical qubit limitations, we investigate quantum error mitigation as a promising alternative.

Quantum error mitigation (QEM) is generally considered in the context of estimating the expectation value of an observable using a quantum circuit, for example in variational quantum eigensolvers \cite{peruzzoVariationalEigenvalueSolver2014} and statistical versions of phase estimation \cite{sommaQuantumEigenvalueEstimation2019, linHeisenbergLimitedGroundStateEnergy2022, wanRandomizedQuantumAlgorithm2022, bluntStatisticalPhaseEstimation2023, gunther2025phase}. QEM methods improve the accuracy of the estimated expectation value by reducing the noise-induced bias in the circuit \cite{temmeErrorMitigationShortDepth2017, caiQuantumErrorMitigation2023}. This is done through post-processing of the measurement outcomes from an ensemble of circuit implementations and so is distinct from quantum error correction which reduces the logical error rate for each individual circuit run \cite{caiQuantumErrorMitigation2023}. Consequently, most error mitigation techniques require additional sampling that increases with the noise in the circuit, resulting in asymptotically unfavourable scaling \cite{piveteauErrorMitigationUniversal2021, caiQuantumErrorMitigation2023}. 

Despite this, QEM techniques could thrive in the early fault-tolerant regime when applied at the logical level by focusing on their pre-asymptotic behaviour prior to the exponential scaling becoming impractical \cite{campbellEarlyFaulttolerantSimulations2022, suzukiQuantumErrorMitigation2022}. As per the standard magic state model \cite{bravyiUniversalQuantumComputation2005}, we assume that logical Clifford operations are ideal; these may be implemented transversely depending on the error-correcting code and underlying hardware. Meanwhile, non-Clifford operations are subject to noise due to imperfect encoded magic states. Given this model, we leverage methods from QEM to simulate single-qubit $Z$-rotation gates in this work, which are an integral part of many quantum algorithms.

QEM ideas have previously been proposed to simulate circuits containing $T$ gates such as in Ref. \cite{piveteauErrorMitigationUniversal2021}. A single-qubit $Z$-rotation gate could then be realized by synthesising into error-corrected Clifford gates and encoded noisy $T$ gates, for example, using the Ross–Selinger method \cite{rossOptimalAncillafreeClifford+T2016}. This leads to a $T$-count (total number of $T$ gates) that scales as $\mathcal{O}(\log_2({1}/{\epsilon_\mathrm{synth}}))$ with accuracy $\epsilon_{\mathrm{synth}}$ of the resultant rotation. The sampling overhead also increases dramatically with the inverse accuracy. This is counter-intuitive as magic resource theory \cite{veitchResourceTheoryStabilizer2014, howardApplicationResourceTheory2017, seddonQuantifyingQuantumSpeedups2021, seddonQuantifyingMagicMultiqubit2019, bravyiSimulationQuantumCircuits2019} shows that a small-angle $Z$-rotation gate is a less powerful resource than a $T$ gate, and yet they are more costly to implement. This perspective hints towards a much more efficient approach to QEM where small-angle $Z$-rotation gates have a low gate and sampling overhead assuming error-corrected Clifford gates with sufficiently low noise levels. This direction has been explored in the Noisy-Intermediate Scale Quantum (NISQ) setting \cite{koczorProbabilisticInterpolationQuantum2024} but not in the context of early fault-tolerant quantum computing.  

In this paper, we introduce a framework that applies the quasiprobability method \cite{howardApplicationResourceTheory2017, pashayanEstimatingOutcomeProbabilities2015} to explore the advantages of decomposing single-qubit $Z$-rotation gates into gates from the Clifford hierarchy \cite{gottesmanDemonstratingViabilityUniversal1999}. A conceptual starting point is to consider the following two-step process: first, use magic state dilution \cite{campbellEfficientMagicState2016} to convert a high magic resource into many low magic resources; and second, use quantum error mitigation to reduce any inherent noise or noise introduced in the dilution process. While each of these two steps could be cast as separate convex optimisation problems, it is more elegant and optimal to compress them. Therefore, rather than implementing these processes independently, our approach unifies them into a single optimisation problem, which we introduce as mitigated magic dilution (MMD). Specifically, we use convex optimisation to find the optimal sample complexity of performing small-angle single-qubit rotations from noisy encoded magic states. 

There has been significant progress in the preparation of encoded magic states. For example, in Ref. \cite{liMagicStatesFidelity2015}, it was shown that encoded magic states can be prepared with a logical error rate of $\sim 0.4 \times 10^{-3}$ under the assumption of ideal single-qubit operations and depolarised two-qubit gates with $0.1\%$ error rate using post-selection. Moreover, there are several improvements to the state preparation of non-stabiliser states for arbitrary small-angle rotations that could offer further advantages to the results presented in this paper, which we examine in \cref{section:discussion}. 

We compare our framework against a baseline classical approach where small-angle single-qubit rotations are decomposed into Clifford operations, similar to Refs. \cite{veitchResourceTheoryStabilizer2014, howardApplicationResourceTheory2017, seddonQuantifyingQuantumSpeedups2021, seddonQuantifyingMagicMultiqubit2019, bravyiSimulationQuantumCircuits2019}. Notably, we demonstrate a polynomial advantage, the magnitude of which depends on the quality of initial magic state preparation. We quantify this advantage in terms of the polynomial degree of magic resource saving of our method. For encoded magic states prepared with 1\% dephasing noise, this saving is better than cubic, while for 0.1\% dephasing noise we find a saving of degree approximately 11.43. Moreover, we show that our method can improve upon a state-of-the-art classical simulator, the sum-over-Cliffords stabilizer extent method \cite{bravyiSimulationQuantumCircuits2019}, offering a 2.37 degree of magic resource saving. 

To evaluate the practical benefits of this approach, we study the resource requirements to simulate the time evolution of the 2D Fermi-Hubbard model. The 2D Fermi-Hubbard model \cite{hubbardElectronCorr1963, HubbardModelHalf2013} is of notable interest in the early fault-tolerant regime \cite{campbellEarlyFaulttolerantSimulations2022, cadeStrategiesSolvingFermiHubbard2020} due to its importance in condensed matter physics (e.g., to understand high-temperature superconductivity \cite{dagottoCorrelatedElectronsHightemperature1994} and the Mott metal-insulator transition \cite{imadaMetalinsulatortransitions}), as well as its simplicity arising from its highly regular lattice structure. Our analysis in \cref{{section:fermi-hubbard}} demonstrates that MMD is a more resource-efficient method than direct gate synthesis, requiring fewer magic states in total. Moreover, the expected number of magic states per sample is significantly smaller, and therefore MMD is particularly amenable to early fault-tolerant devices. 

Even when quantum computers have an asymptotic speedup over classical computing they can fail to have an in-practice speed-up for example problems of a relevant size \cite{babbushBeyondQuadraticSpeedups2021}, and so it is crucial to make such comparison. Here, we consider the example problem of a Fermi-Hubbard model with a $6 \times 6$ square lattice and evolution time $t=0.25$ as a strong candidate for quantum advantage. We find that our MMD method requires a circuit with only $1037$ non-Clifford gate teleportations (on average) and $5.34\times10^{6}$ samples. Even with error correction overheads, a sample per second is a conservative runtime estimate, taking a single quantum computer $62$ days for all samples. For the same calculation, the sum-over-Cliffords stabilizer extent method would require a $5.18 \times 10^{17}$ seconds runtime, which would take $1.64 \times 10^{4}$ years to complete when assuming a million classical processors in parallel. Thereby, our MMD protocol facilitates a significant reduction in runtime over the the sum-over-Cliffords classical simulator for conservative time evolution and lattice sizes of interest. Combined with a Pauli-based model of computation \cite{litinskiGameSurfaceCodes2019}, such an application would require logical error rates of approximately 1 part in a millon, the so-called MegaQuOp regime \cite{preskillBeyondNISQMegaquop2025}.

This paper is structured as follows. In \cref{sec:QPM} we briefly summarise the quasiprobability method. In \cref{sec:methods}, we introduce our framework in two steps, working in the channel representation of all unitary gates $U$ throughout, which are denoted as $\mathcal{U}(\cdot) = U(\cdot)U^{\dag}$. We first present the application of the quasiprobability method to target $\mathcal{Z}$ rotation channels over diagonal Clifford + $\mathcal{T}$ channels in \cref{subsec:LCC} using a linear combination of channels decomposition. We then generalise this to decompositions over diagonal Clifford + $\mathcal{T}^\frac{1}{n}$ channels in \cref{subsec:hierarchy}. Finally, in \cref{section:fermi-hubbard}, we demonstrate the practical applicability of our framework for a second-order Trotter simulation of the Fermi-Hubbard model with comparison to gate synthesis.

\section{\label{sec:QPM}Quasiprobability method}
The motivation behind the quasiprobability method is that an ideal quantum operation can be decomposed into a basis set of noisy operations. Let $\mathcal{U}^{\text{t}}(\rho) = U^{\text{t}} \rho {U^{\text{t}}}^{\dag} $ be the ideal target unitary channel of a quantum operation $U^{\text{t}}$ and let $\{\mathcal{U}^{\text{n}}_i\}$ be a set of channels corresponding to the noisy operations that can be performed on a given quantum hardware. The noise on these operations would typically be characterised using a tomography procedure \cite{endoPracticalQuantumError2018, caiQuantumErrorMitigation2023}. We can then write the following decomposition, hereafter referred to as the linear combinations of channels (LCC) decomposition
\begin{equation}
    \mathcal{U}^{\text{t}}= \sum_i x_i \mathcal{U}^{\text{n}}_i
    \label{eqn:qpm}
\end{equation}
where $x_i$ are real coefficients that can be positive or negative such that \cref{eqn:qpm} is a quasiprobability representation.

It follows that the expectation value of an observable $O$ for the target operation can be written in terms of the expectation value of the noisy operations as
\begin{equation}
\begin{split}
     \text{Tr}\big[O\mathcal{U}^{\text{t}}(\rho)\big] &= 
     \text{Tr}\big[O\sum_i x_i \mathcal{U}^{\text{n}}_i(\rho)\big] \\
     &= \sum_i x_i\text{Tr}\big[O \mathcal{U}^{\text{n}}_i(\rho)\big]. 
    \label{eqn:qpm-2}
\end{split}
\end{equation}

The expectation values $\text{Tr}\big[O \mathcal{U}^{\text{n}}_i(\rho)\big]$ associated with noisy channels $\mathcal{U}^{\text{n}}_i$ can be estimated using Monte Carlo sampling. Each noisy operation $\mathcal{U}^{\text{n}}_i$, specified by index $i$, is applied with probability ${|x_i|}/{\sum_i |x_i|}$, and the resulting expectation value is multiplied by $\text{sign}(x_i) \sum_i |x_i|$ \cite{howardApplicationResourceTheory2017, caiQuantumErrorMitigation2023}. From this, an estimate of the ideal expectation value can be calculated up to an accuracy $\epsilon$ and probability greater than $1 - \delta$, where
\begin{equation}
    \delta = 2 \exp\bigg({\frac{-N\epsilon^2}{2 (\sum_i |x_i|)^2}}\bigg)
\end{equation}
and the number of samples $N$ is determined by Hoeffding's inequality to be 
\begin{equation}
\label{eq:samples}
    N = \frac{2}{\epsilon^2} \bigg( \sum_i |x_i| \bigg)^2 \ln \bigg( \frac{2}{\delta} \bigg).
\end{equation}

Thus, the number of samples scale as $\mathcal{O}\big({\lambda^2}/{\epsilon^2}\big)$ up to logarithmic factors, with $\lambda = \sum_i |x_i|$. The quantity $\lambda^2$ is referred to as the \textit{sampling overhead} of using this method over finding the expectation value of the ideal target channel directly \cite{piveteauErrorMitigationUniversal2021, caiQuantumErrorMitigation2023}.

\section{\label{sec:methods}Methodology}

\begin{figure}
    \centering
    \includegraphics[width=0.8\linewidth]{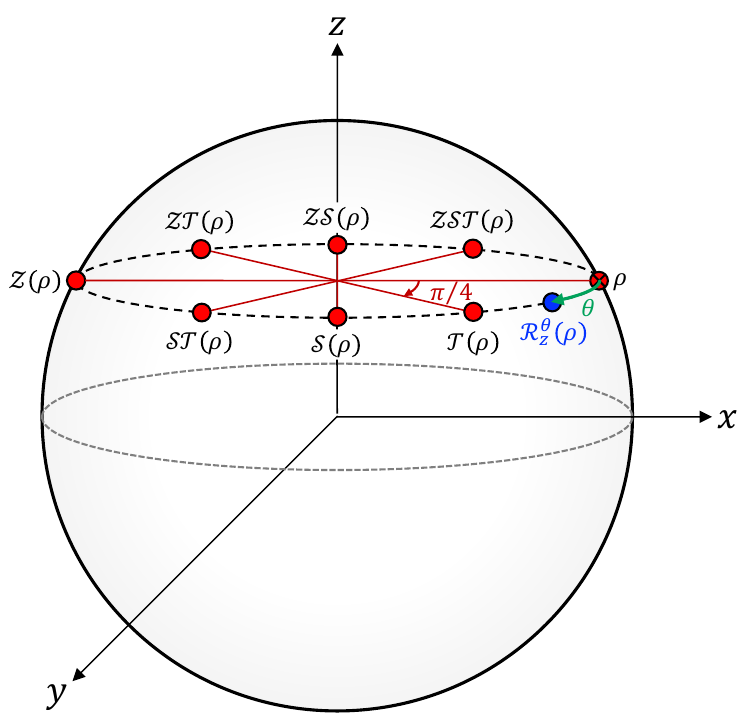}
    \caption{\label{fig:1}Action of $\mathcal{T}^{k}$ rotation channels (where $1 \leq k \leq 8$) on a density matrix $\rho$. The eight operations (in red) form the initial basis set $\mathcal{G}$. The target rotation channel $\mathcal{R}_z^{\theta}(\rho)$ is shown in blue.}
\end{figure}

\subsection{\label{subsec:LCC}Linear Combination of Channels (LCC) Decomposition}
We first demonstrate how the quasiprobability method can be used to decompose target $\mathcal{Z}$-rotation channels. In the simplest case, we consider a diagonal Cliffords + $\mathcal{T}$ channel decomposition, forming an overcomplete basis set to ensure that we can optimise this decomposition with respect to the sampling overhead.

Consider the action of a small-angle $\mathcal{Z}$-rotation channel $\mathcal{R}^{\theta}_z(\rho) = R_z(\theta)\rho(R_z(\theta))^{\dag} $ acting on a state represented by density matrix $\rho$ in \cref{fig:1}, where $R_z(\theta) = \exp (- i (\theta/2) Z)$ is a gate representing a rotation about the $Z$-axis with angle $\theta$. In this geometric picture, we label eight channels corresponding to the action of a basis set of gates 
\begin{equation}
    \begin{split}
    G &= \{T^k : 1 \leq k \leq 8 \} \\ &= \{I, T, S, ST, Z, ZT, ZS, ZST\},
    \end{split}
    \label{eqn:basis set of gates}
\end{equation}
which forms a cyclic group $(G, \times)$ generated by the $T$ gate under multiplication. We represent these gates in their channel representation to form a set $\mathcal{G}$ such that 
\begin{equation}
    \mathcal{G} = \{\mathcal{T}^k(\rho) = (T^k)\rho(T^k)^\dag : 1 \leq k \leq 8 \}.
    \label{eqn:basis set of channels}
\end{equation}
The LCC decomposition is given by 
\begin{equation}
    \mathcal{R}^\theta_z(\cdot) = \sum_{U \in G} x_U U (\cdot) U^\dag = \sum_{\mathcal{U} \in \mathcal{G}} x_U\mathcal{U}(\cdot),
    \label{eqn:lcc decomposition}
\end{equation}
where $\mathcal{R}^\theta_z(\cdot)$ represents the single qubit $\mathcal{Z}$-rotation channel for the $R_z(\theta)$ gate. This is a quasiprobability decomposition such that the coefficients $x_U$ are real and satisfy $\sum_{\mathcal{U} \in \mathcal{G}} x_U = 1$. The quantity $\lambda$ is defined as the $l_1$-norm of the coefficients in this decomposition, that is 
\begin{equation}
    \lambda = \lVert x \rVert_1 = \sum_{U \in G} |x_U|, 
\end{equation}
such that $\lambda^2$ is the sampling overhead of each decomposition from Hoeffding's inequality \cite{piveteauErrorMitigationUniversal2021,temmeErrorMitigationShortDepth2017}. 

It can be noted that the decomposition in \cref{eqn:lcc decomposition} is not unique over a chosen basis set and so it is important to minimise the sampling overhead over different solutions. We define the optimal decomposition for a given $\mathcal{Z}$-rotation channel as the decomposition into elements of $\mathcal{G}$ with the minimum $\lambda$, which we denote as $\Lambda_\mathcal{G}(\mathcal{R}^\theta_z)$. This can be written as 
\begin{equation}
    \Lambda_\mathcal{G}(\mathcal{R}^\theta_z(\cdot)) = \min \bigg\{ \lambda = \lVert x_U \rVert_1 \bigg| \mathcal{R}^\theta_z(\cdot) = \sum_{\mathcal{U} \in \mathcal{G}} x_U \mathcal{U} (\cdot) \bigg\}.
    \label{eqn:defn of Lambda}
\end{equation}
By definition, $\Lambda_\mathcal{G}(\mathcal{R}^\theta_z) \geq 1$ where $\Lambda_\mathcal{G}(\mathcal{R}^\theta_z) = 1$ if $\mathcal{R}^\theta_z \in \mathcal{G}$ for some group $\mathcal{G}$ representing the basis set of channels. The solution to \cref{eqn:defn of Lambda} is equivalent to minimising the sampling overhead $\lambda^2$ over the basis set of channels, thus $\Lambda_\mathcal{G}(\mathcal{R}^\theta_z(\cdot))$ finds the optimal decomposition of $\mathcal{R}^\theta_z(\cdot)$ for the chosen group $\mathcal{G}$. This can be further expressed as a convex optimisation problem to solve a linear system given by
\begin{equation}
    \Lambda_\mathcal{G}(\mathcal{R}^\theta_z(\cdot)) = \min \lVert x \rVert_1 \text{ subject to } Ax = b.
    \label{eqn:convex}
\end{equation}
Here, vector $b$ represents $m$ elements of the target channel $\mathcal{R}^\theta_z$ expressed in a vectorised form, and $A$ is an $m \times n$ matrix, with $n$ columns for the basis set of channels in the decomposition and $m$ rows of elements that specify each channel (see \cref{eqn:MatA} and \cref{eqn:Vecb} as an example). The $j^{\text{th}}$ column of matrix $A$ can be generated by determining the vectorised form of the $j^{\text{th}}$ channel $\mathcal{U}_j(\cdot) = U_j (\cdot) U_j^{\dag}$ for $\mathcal{U}_j \in \mathcal{G}$ and $j \in \{1, 2, \dots, n\}$. Where possible, we also make a concerted effort to minimise the number of non-zero elements in $x$ whilst keeping the optimal sampling overhead constant. 

For the ideal basis set of channels $\mathcal{G}$ defined in \cref{eqn:basis set of channels}, we use \textsc{CVXPY} \cite{diamond2016cvxpy, agrawal2018rewriting} to solve \cref{eqn:convex} and find that decompositions of the form
\begin{equation}
    \mathcal{R}^\theta_z(\cdot) = x_I\mathcal{I}(\cdot) + x_T\mathcal{T}(\cdot) + x_Z\mathcal{Z}(\cdot)
    \label{optimal-LCC}
\end{equation}
are optimal, with
\begin{equation}
    \Lambda_\mathcal{G}(\mathcal{R}^\theta_z(\cdot)) = (\sqrt{2} - 1)\sin{(\theta)} + \cos{(\theta)}
    \label{eqn:Lambda_G}
\end{equation}
for $0 \leq \theta \leq \frac{\pi}{4}$. Additionally, considering subsets of $\mathcal{G}$ with only 2 elements, we find that a solution to the optimisation problem cannot be found, and therefore 3 contributions are required (see \cref{app:A}).

As a classical baseline,  we consider the subset of $\mathcal{G}$ that are Clifford operations, denoted as $\mathcal{C}$. In that case, the optimal decomposition consists of $\{\mathcal{I}, \mathcal{S}, \mathcal{Z}\}$ channels and the corresponding $l_1$-norm $\Lambda_\mathcal{C}$ is 
\begin{equation}
    \Lambda_\mathcal{C}(\mathcal{R}^\theta_z(\cdot)) = \sin{(\theta)} + \cos{(\theta)}.
    \label{eqn:Lambda_C}
\end{equation}
We compare this baseline to our quantum protocol by considering $(\Lambda_\mathcal{G}^2)^\gamma = \Lambda_\mathcal{C}^{2}$, which implicitly defines $\gamma$ where
\begin{equation}
    \gamma(\theta) = \frac{\ln(\Lambda_\mathcal{C}(\mathcal{R}_z^\theta))}{\ln(\Lambda_\mathcal{G}(\mathcal{R}_z^\theta))}.
    \label{eqn:gamma defn}
\end{equation}
Consequently, we interpret $\gamma$ as the polynomial degree of magic resource saving of the optimal decomposition over $\mathcal{G}$ compared to the optimal decomposition over $\mathcal{C}$, for a specified target rotation channel. As $\theta \to 0$, $\gamma$ approaches $\sqrt{2}+1 \approx 2.41 > 2$, indicating a slightly better-than-quadratic advantage of decomposing a (very) small-angle rotation into a quasiprobability decomposition of $\{\mathcal{I}, \mathcal{T}, \mathcal{Z}\}$ channels instead of $\{\mathcal{I}, \mathcal{S}, \mathcal{Z}\}$ channels. From this, we can now proceed to generalise these findings to achieve a better advantage for small-angle rotations by our choice of $\mathcal{G}$ from the Clifford hierarchy. 

\subsection{\label{subsec:hierarchy}Climbing the Clifford Hierarchy}

From \cref{eqn:Lambda_G} and \cref{eqn:Lambda_C}, it can be seen that the optimal channel decompositions over our choice of $\mathcal{G}$ and $\mathcal{C}$ respectively are of the form $\{\mathcal{I}, \mathcal{R}_z^{\phi}, \mathcal{Z} \}$. Specifically, $\phi = \pi/2$ for $\mathcal{R}_z^{\phi} = \mathcal{S}$ and  $\phi = \pi/4$ for $\mathcal{R}_z^{\phi} = \mathcal{T}$ with optimality valid within the range $0 \leq \theta \leq \phi$. Therefore, for smaller target rotations, we can minimise $\Lambda_\mathcal{G}(\mathcal{R}^\theta_z(\cdot))$ further by including $\mathcal{R}_z^{\phi}$ channels with smaller $\phi$ in our choice of $\mathcal{G}$. 

\begin{figure}[h]
    \centering
    \begin{quantikz}
    \lstick{\ket{\psi}}                             & \ctrl{4} & \ctrl{2} \gategroup[3, steps=3, style=dashed]{} & \ctrl{1}  \gategroup[2, steps=2, style=dashed]{} & \gate[style=dashed]{S} & & \rstick{$\sqrt[n]{T}$\ket{\psi}} \\
    \lstick{$|T\rangle$ = $T$\ket{+}}               & & & \targ{} & \meter{} \wire[u][1]{c}\\
    \lstick{$|\sqrt{T}\rangle$ = $\sqrt{T}$\ket{+}}  & & \targ{} & & \metercw{} \\ 
    \vdots &\wireoverride{n} &\wireoverride{n} &\wireoverride{n} &\wireoverride{n} \vdots \\
    \lstick{$|\sqrt[n]{T}\rangle$ = $\sqrt[n]{T}$\ket{+}} & \targ{} & & & \metercw{} 
    \end{quantikz}
    \caption{\label{fig:generalised circuit}Generalised teleportation circuit to implement a $\sqrt[n]{T}$ gate using $i$ distinct magic states, where $n = 2^{i-1}$, and Clifford operations. Boxes and gates with a dashed line are classically controlled; they are only implemented if the measurement below obtains an outcome with eigenvalue $-1$.}
\end{figure}
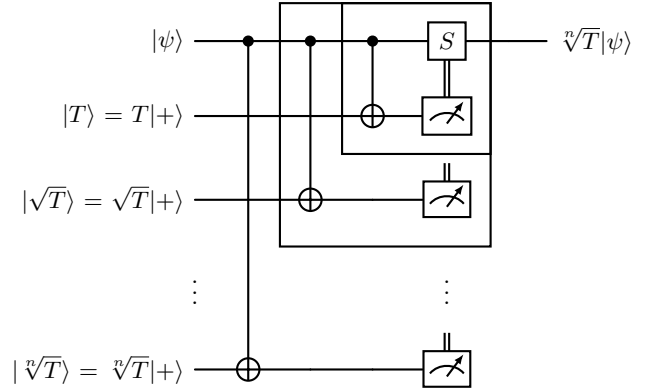

First, we let $i \in \mathbb{Z_{\geq \text{0}}}$ and define $n = 2^{i-1}$. We then choose $\mathcal{G}$ to be 
\begin{equation}
    \mathcal{G} = \{\mathcal{T}^{\frac{k}{n}}(\rho) = (T^{\frac{k}{n}})\rho(T^{\frac{k}{n}})^\dag : 1 \leq k \leq 8n \},
    \label{eqn:G with n-th root T}
\end{equation}
where $T^{\frac{1}{n}}$ is an $n^{\text{th}}$-root $T$ gate from the $(i+2)$-th level of the Clifford hierarchy, and $\phi = \pi / 4n$. Therefore, $i=0$ represents the $S$ gate, $i=1$ represents the $T$ gate, and so on. These $T^{\frac{1}{n}}$ gates can be implemented by a generalised gate teleportation circuit as shown in \cref{fig:generalised circuit}.

We find that the LCC decomposition for \cref{eqn:convex} with $\mathcal{G}$ as \cref{eqn:G with n-th root T} that minimises the sampling overhead is given by
\begin{equation}
    \mathcal{R}_z^{\theta}(\cdot) = x_I \mathcal{I}(\cdot) + x_n \mathcal{T}^{\frac{1}{n}}(\cdot) + x_Z \mathcal{Z}(\cdot),
    \label{optimal-LCC-n}
\end{equation} 
with corresponding $\Lambda_\mathcal{G}(\mathcal{R}^\theta_z(\cdot))$ as 
\begin{equation}
    \Lambda_\mathcal{G}(\mathcal{R}^\theta_z(\cdot)) = \cos(\theta) + \frac{\sin(\theta)}{\sin(\phi)} \bigg(1 -\cos(\phi) \bigg).
    \label{eqn:ideal phi}
\end{equation}

For small-angle rotations, $\Lambda_\mathcal{G}(\mathcal{R}^\theta_z(\cdot))$ can be approximated to be 
\begin{equation}
    \Lambda_\mathcal{G}(\mathcal{R}^\theta_z(\cdot)) \approx 1 + \theta \bigg( \frac{1 - \cos(\phi)}{\sin(\phi)}\bigg), 
\end{equation}
with a degree of resource saving compared to a Clifford decomposition of
\begin{equation}
   \gamma = \frac{\sin(\phi)}{1 - \cos(\phi)} = \cot\bigg({\frac{\phi}{2}}\bigg)
\end{equation}
in the limit of small $\theta$ (see \cref{app:B1}).
Therefore in the $\{\mathcal{I}, \mathcal{T}^\frac{1}{n}, \mathcal{Z}\}$ decomposition, climbing the Clifford hierarchy in terms of $\mathcal{T}^\frac{1}{n}$ channels  results in a larger degree of saving compared to the $\{\mathcal{I}, \mathcal{S}, \mathcal{Z}\}$ Clifford decomposition. 

We now consider the case where the non-Clifford channels in our basis set are subject to noise, as originally motivated in the quasiprobability method. Non-Clifford operations (in our case $\mathcal{T}^{\frac{1}{n}}$) can be implemented via encoded gate teleportation circuits using noisy encoded magic states and error-corrected Clifford gates as shown in \cref{fig:generalised circuit}. The most relevant noise channel for these diagonal rotation gates in our basis set is the dephasing noise channel (see \cref{app:C}), which is defined as
\begin{equation}
    \varepsilon(\cdot) = (1 - p)(\cdot) + pZ(\cdot)Z,
\end{equation}
where $p$ quantifies the amount of dephasing noise. The action of the dephasing noise channel on a unitary channel in our basis set of unitary operations is given as
\begin{equation}
   \mathcal{U}^{\text{deph}}(\cdot) = \varepsilon(\mathcal{U}^{\text{ideal}}(\cdot)) = (1 - p)\mathcal{U}^{\text{ideal}}(\cdot)  + p \mathcal{Z} \circ \mathcal{U}^{\text{ideal}}(\cdot).
   \label{eqn:dephased U(p)}
\end{equation}
As the number of iterations of the teleportation circuit increases for the implementation of gates from higher levels of the Clifford hierarchy (\cref{fig:generalised circuit}), we find that the dephasing noise changes correspondingly. We assume that each non-stabiliser state $|T^{\frac{1}{n}}\rangle$ is prepared with uniform  (independent of $\theta$) fidelity. However, in \cref{section:discussion}, we discuss how $\theta$-dependent fidelities could lead to further performance improvements. 

If the dephasing noise for a $T$ gate implementation (requiring a $|T\rangle$ magic state) is $p$, the effective dephasing noise for a $T^\frac{1}{n}$ gate can be found by considering the errors on the non-stabiliser states that lead to an error on the $T^\frac{1}{n}$ gate. For example, the probability of a $Z$ error acting on a $T^\frac{1}{2}$ gate is given by the probability of an error occurring on either the $|T\rangle$ or $|T^\frac{1}{2}\rangle$ state \footnote{Note that if $Z$ errors occur on both the $|T\rangle$ and $|T^\frac{1}{2}\rangle$ states, then the overall effect on the $T^\frac{1}{2}$ gate is $Z^2=I$, meaning this case is equivalent to no error occurring.}, that is 
\begin{equation}
\begin{split}
p_{\text{eff}} &= p\bigg(1-\frac{p}{2}\bigg) + \frac{p}{2}(1-p) = \frac{3}{2}p - p^2 < \frac{3}{2}p,
\end{split}
\end{equation}
where we take into account that there is a 50\% probability that a $T$ gate correction needs to be applied (recall \cref{fig:generalised circuit}). Therefore, the effective dephasing noise on a $T^\frac{1}{n}$ gate is bounded by $p_{\text{eff}} = \left( 2 - n^{-1} \right) p$.

As a result of this, the $l_1$-norm for the $\{\mathcal{I}, \varepsilon(\mathcal{T}^\frac{1}n{}), \mathcal{Z}\}$ decomposition transforms from the ideal case in \cref{eqn:ideal phi} to the following:
\begin{widetext}
\begin{equation}
\begin{split}
     \Lambda_\mathcal{G}(\mathcal{R}^\theta_z(\cdot)) 
    =& \bigg|  \cos^2{\bigg(\frac{\theta}{2}\bigg)} - \bigg[ \cos^2{\bigg(\frac{\phi}{2}\bigg)}  - p_{\text{eff}}\cos{(\phi)} \bigg]\frac{\sin(\theta)}{(1-2p_{\text{eff}})\sin(\phi)}\bigg| 
    \\ &+ \bigg| \frac{\sin(\theta)}{(1-2p_{\text{eff}})\sin(\phi)}  \bigg| 
    \\ &+ \bigg| \sin^2{\bigg(\frac{\theta}{2}\bigg)} - \bigg[ \sin^2{\bigg(\frac{\phi}{2}\bigg)} + p_{\text{eff}} \cos{(\phi)}\bigg]\frac{\sin(\theta)}{(1-2p_{\text{eff}})\sin(\phi)}\bigg|.
\end{split}
\end{equation}
\end{widetext}

\begin{figure}[t!]
    \centering
    \includegraphics[width=\linewidth]{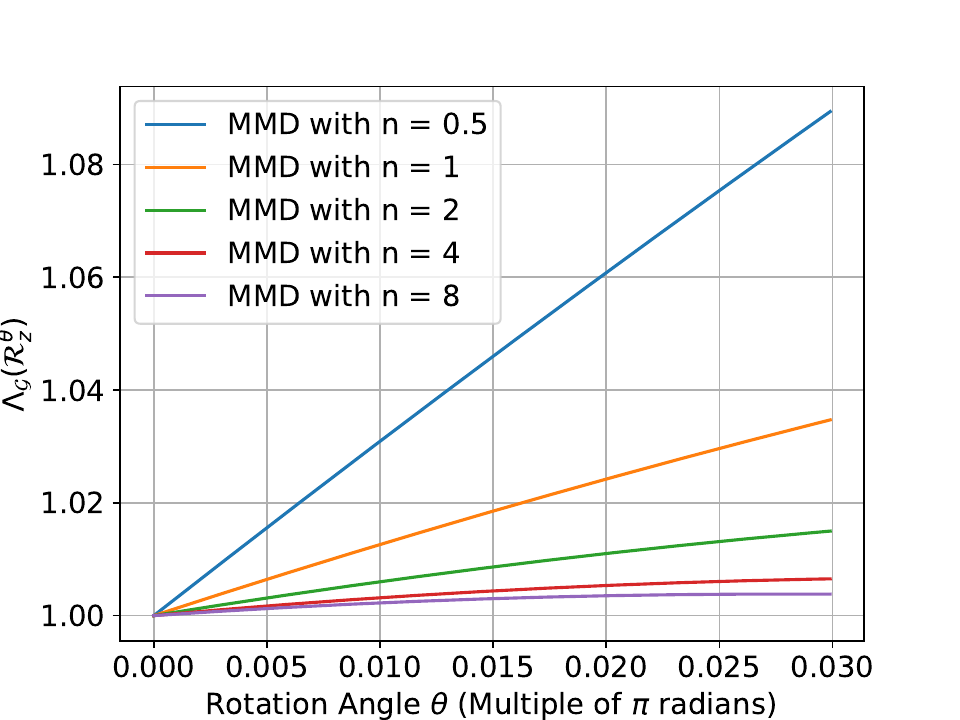}
    \caption{\label{fig:lambda-vs-theta}$\Lambda_{G}(\mathcal{R}_z^{\theta})$ as a function of target rotation angle $\theta$ for different optimal decompositions of the form $\{\mathcal{I}, \varepsilon(\mathcal{T}^\frac{1}n{}), \mathcal{Z}\}$ including the optimal Clifford decomposition of $\{\mathcal{I}, \mathcal{S}, \mathcal{Z}\}$. A dephasing error of $0.1\%$ is assumed for non-Clifford state preparation.} 
\end{figure}

In \cref{fig:lambda-vs-theta} we present $\Lambda_\mathcal{G}(\mathcal{R}^\theta_z)$ as function of $\theta$ for $\{\mathcal{I}, \varepsilon(\mathcal{T}^\frac{1}n{}), \mathcal{Z}\}$ decompositions with $n \in \{0.5, 1, 2, 4, 8\}$ for $p=0.1\%$ dephasing noise, where $n=0.5$ corresponds to the $\mathcal{S}$ channel. 

In particular, we are interested in the degree of magic resource saving $\gamma$ to benchmark against the classical Clifford decomposition. As $\theta \to 0$, $\gamma$ with respect to the $\{ \mathcal{I}, \mathcal{S}, \mathcal{Z}\}$ Clifford decomposition is approximated by (see \cref{app:B2})
\begin{equation}
    \frac{1}{\gamma} \approx \frac{\csc(\phi)}{(1-2p_{\text{eff}})} - \cot(\phi).
    \label{eqn:relative overhead dephased R_phi}
\end{equation}

\begin{figure}[h]
    \centering
    \includegraphics[width=\linewidth]{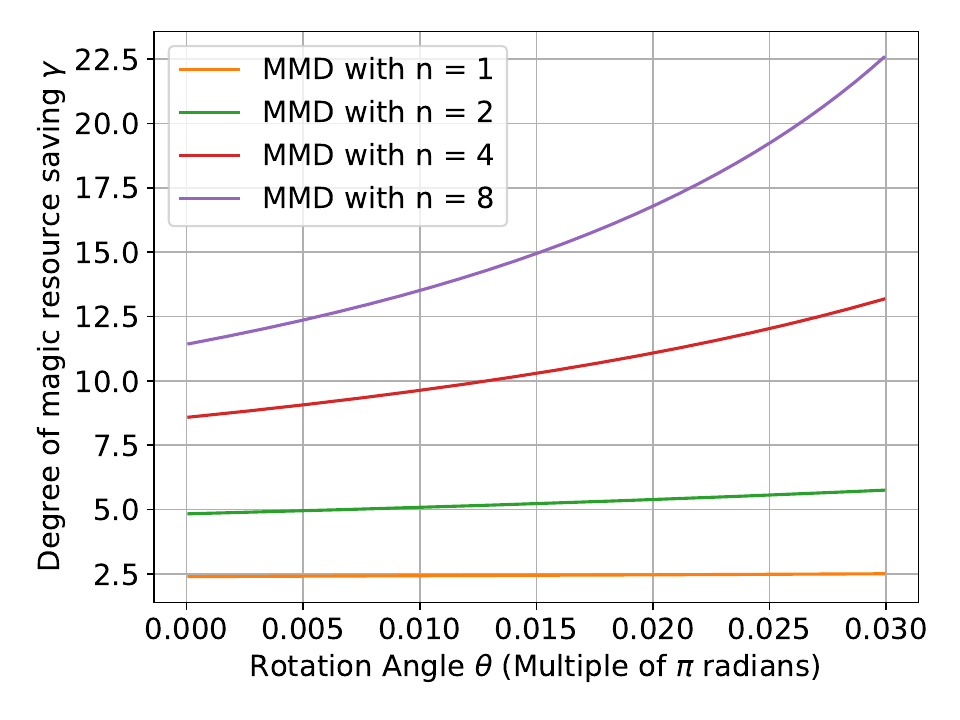}
    \caption{\label{fig:gamma-vs-theta}Degree of magic resource saving $\gamma$ as a function of target rotation angle $\theta$ for different optimal decompositions of the form $\{\mathcal{I}, \varepsilon(\mathcal{T}^\frac{1}{n}), \mathcal{Z}\}$ relative to the optimal Clifford decomposition of $\{\mathcal{I}, \mathcal{S}, \mathcal{Z}\}$. A dephasing error of $0.1\%$ is assumed for non-Clifford state preparation.} 
\end{figure}

From \cref{fig:gamma-vs-theta}, we find that for $p=0.1\%$ and in the limit of small $\theta$, we can achieve at least a $\sim 2.4$ degree of saving in the simplest case of a $\{\mathcal{I}, \varepsilon(\mathcal{T}), \mathcal{Z}\}$ decomposition, and greater than $\sim 11.4$ degree of saving relative to the Clifford decomposition by replacing $\mathcal{T}$ with a $\mathcal{T}^\frac{1}{8}$ channel. 

We further note the dependence of $\Lambda_{G}(\mathcal{R}_z^{\theta})$ on the dephasing noise $p$. For $p=0.1\%$, increasing $n$ in our protocol results in a greater advantage up to $n=8$, as shown in \cref{fig:gamma-vs-theta}. However, for higher $p$, this is not always the case. In fact, for $1\%$ dephasing noise, going beyond $n=2$ presents no further improvement in resource saving as shown in \cref{table:1} (see \cref{app:D} for corresponding values of $\ln(\Lambda_{\mathcal{G}}(\mathcal{R}_z^{\theta}))$). In other words, the $\{\mathcal{I},\varepsilon(\mathcal{T}^\frac{1}{2}),\mathcal{Z}\}$ decomposition is optimal and provides a lower bound of $\sim 3.6$ degree of saving for small $\theta$. In general, the smallest value of $n$ that provides the optimal saving can be found by maximising $\gamma$ with respect to $n$ (indicated in bold in \cref{table:1}).

\begin{table}[h]
\caption{\label{table:1}Degree of magic resource saving $\gamma$ for a basis set of channels $\mathcal{G}$ with optimal channel decomposition $\{\mathcal{I}, \varepsilon(\mathcal{T}^\frac{1}{n}), \mathcal{Z}\}$ relative to the $\{\mathcal{I}, \mathcal{S}, \mathcal{Z}\}$ channel combination for increasing $n$ up to the sixth level of the Clifford hierarchy, with non-Clifford gates subject to dephasing noise with probability $p$.}
\centering
\begin{tabular}{|c|c|c|c|c|c|c|}
\hline
$n$ & $\phi$ & $\mathcal{R}_z^{\phi}$ & \multicolumn{4}{|c|}{$\gamma$ (Small $\theta$)} \\
\hline
& & & $p = 0.01\%$ & $p = 0.1\%$ & $p = 0.5\%$ & $p = 1.0\%$ \\
\hline
1 & $\frac{\pi}{4}$ & $\mathcal{T}$ & 2.41 & 2.40 & 2.33 & 2.26  \\
2 & $\frac{\pi}{8}$ & $\mathcal{\sqrt{T}}$ & 5.01 & 4.84 & 4.19 & \textbf{3.58} \\
4 & $\frac{\pi}{16}$ & $\mathcal{\sqrt[4]{T}}$ & 9.97 & 8.58 & \textbf{5.27} & 3.52 \\
8 & $\frac{\pi}{32}$ & $\mathcal{\sqrt[8]{T}}$ &18.88 &\textbf{11.43} & 4.10 & 2.24\\
\hline
\end{tabular}
\end{table}

\begin{figure}[t!]
    \centering
    \includegraphics[width=\linewidth]{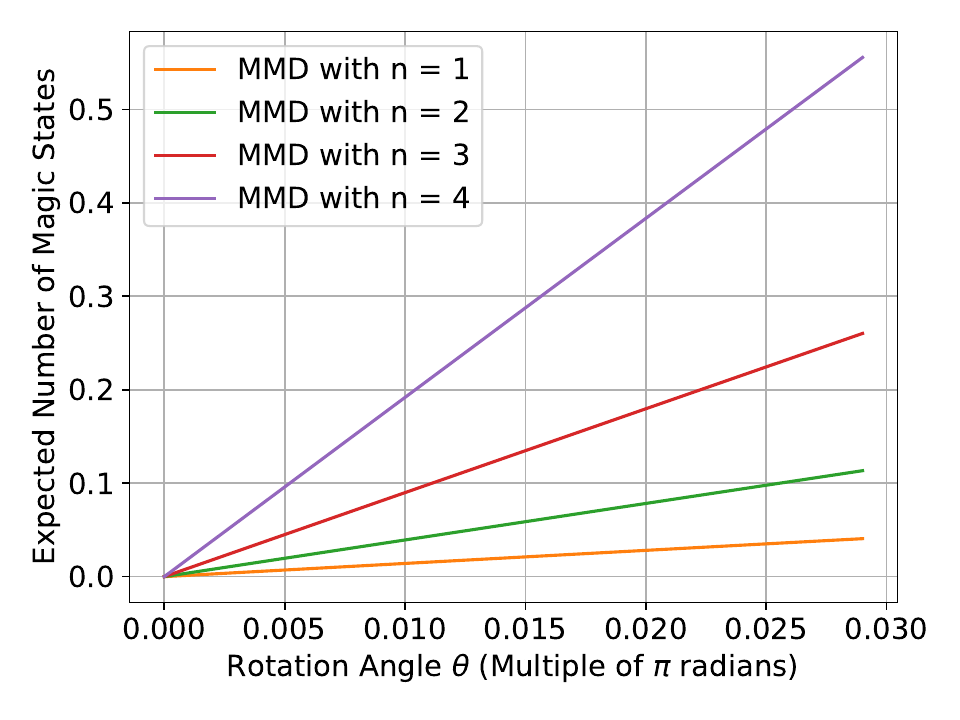}
    \caption{\label{fig:proportion magic states}Expected number of magic states $E$ per sample as a function of target rotation angle $\theta$ for different optimal decompositions of the form $\{\mathcal{I}, \varepsilon(\mathcal{T}^\frac{1}{n}), \mathcal{Z}\}$ relative to the optimal Clifford decomposition of $\{\mathcal{I}, \mathcal{S}, \mathcal{Z}\}$. A dephasing error of $0.1\%$ is assumed for non-Clifford state preparation.} 
\end{figure}

A further useful metric to consider is the expected number of magic states required per sample for an optimal decomposition of the form \cref{optimal-LCC-n} which we derive from the proportion of $|x_n|$ relative to $\Lambda_\mathcal{G}(\mathcal{R}^{\theta}_z)$, and the total number of magic states (per sample) required to implement a $T^\frac{1}{n}$ gate using \cref{fig:generalised circuit}:

\begin{equation}
    E = \bigg(2 - \frac{1}{n}\bigg) \frac{\big|x_n\big|}{\Lambda_\mathcal{G}(\mathcal{R}^{\theta}_z)}.
    \label{eqn:expected-magic-states}
\end{equation} 
In this expression, $x_n$ is the coefficient of the $\mathcal{T}^\frac{1}{n}$ term in the LCC decomposition of the form $\{\mathcal{I}, \varepsilon(\mathcal{T}^\frac{1}n{}), \mathcal{Z}\}$, and $\Lambda_\mathcal{G}$ is the $l_1$-norm of the coefficients in this decomposition. As the target rotation angle $\theta$ decreases, $|x_n|$ also decreases, resulting in a smaller number of expected magic states. Thus, this method ``dilutes'' magic resource in simulating smaller rotations from $\mathcal{T}^\frac{1}{n}$ channels of higher magic resource. This dilution phenomenon is evident in \cref{fig:proportion magic states}, which shows the expected number of magic states (per sample) decreasing with the rotation angle $\theta$.

We also present a comparison of our quantum protocol to the classical sum-over-Cliffords simulation method \cite{bravyiSimulationQuantumCircuits2019}. The total runtime of this method scales as $\mathcal{O}(\xi/\epsilon^4)$ where $\xi$ is the stabilizer extent and $\epsilon$ is the precision of the simulation. Therefore, the sum-over-Cliffords method has a worse scaling w.r.t. the desired precision than MMD where the scaling is proportional to $1/\epsilon^2$ as in \cref{eq:samples}. In the small $\theta$ limit, the stabilizer extent for a $Z$-rotation is given by 
\begin{equation}
    \xi(R(\theta)) = \exp(\tan(\pi/8)\theta).
\end{equation}
Therefore, we can define a resource saving degree $\gamma_{\text{\tiny{SE}}}$ such that $(\Lambda_\mathcal{G}^2)^{\gamma_{\text{\tiny{SE}}}} = \xi$ analogously to \cref{eqn:relative overhead dephased R_phi}. This results in 
\begin{equation}
     {\gamma_{\text{\tiny{SE}}}}(\theta) = \frac{\ln(\xi(R(\theta)))}{2\ln(\Lambda_\mathcal{G}(\mathcal{R}_z^\theta))}.
\end{equation}
We present this with the caveat that MMD has an additional advantage with respect to precision $\epsilon$ scaling, which is not captured by $\gamma_\mathrm{SE}$. 

\begin{table}[h!]
\caption{\label{table:2}Degree of magic resource saving $\gamma_{\text{\tiny{SE}}}$ for a basis set of channels $\mathcal{G}$ with optimal channel decomposition $\{\mathcal{I}, \varepsilon(\mathcal{T}^\frac{1}n{}), \mathcal{Z}\}$ relative to the stabiliser extent method for increasing $n$ up to the sixth level of the Clifford hierarchy, with non-Clifford gates subject to dephasing noise with probability $p$.}
\centering
\begin{tabular}{|c|c|c|c|c|c|c|}
\hline
$n$ & $\phi$ & $\mathcal{R}_z^{\phi}$ & \multicolumn{4}{|c|}{$\gamma_{\text{\tiny{SE}}}$ (Small $\theta$)} \\
\hline
& & & $p = 0.01\%$ & $p = 0.1\%$ & $p = 0.5\%$ & $p = 1.0\%$ \\
\hline
1 & $\frac{\pi}{4}$ & $\mathcal{T}$ & 0.50 & 0.50 & 0.48 & 0.47  \\
2 & $\frac{\pi}{8}$ & $\mathcal{\sqrt{T}}$ & 1.04 & 1.00 & 0.87 & \textbf{0.74} \\
4 & $\frac{\pi}{16}$ & $\mathcal{\sqrt[4]{T}}$ & 2.07 & 1.78 & \textbf{1.09} & 0.73 \\
8 & $\frac{\pi}{32}$ & $\mathcal{\sqrt[8]{T}}$ & 3.91 &\textbf{2.37} & 0.85 & 0.46\\
\hline
\end{tabular}
\end{table}

As shown in \cref{table:2}, the MMD method is able to achieve a better-than-quadratic advantage for $p=0.1\%$ dephasing noise with respect to the stabilizer extent. Meanwhile for higher dephasing noise, our method provides comparable performance with a slight increase in magic resource for $p=1.0\%$. 

\section{\label{section:fermi-hubbard}Fermi-Hubbard Model Simulation}

The Fermi-Hubbard model describes the behaviour of interacting electrons in 2D materials. The Hubbard Hamiltonian is composed of hopping terms $H_h$ which represent the kinetic energy of electrons that can tunnel between neighbouring lattice sites, and interaction terms $H_i$ which represent the potential energy due to the on-site repulsion of electrons. For this analysis, we consider $H_i$ subject to a chemical potential shift as per Ref. \cite{campbellEarlyFaulttolerantSimulations2022}. The resulting Hamiltonian takes the form 
\begin{equation}
    \begin{split}
    H &= H_h + H_i \\
    &= \sum_{\sigma \in \uparrow, \downarrow} \sum_{i \neq j} R_{i, j} a^{\dag}_{i, \sigma} a_{j, \sigma} + \frac{u}{4} \sum_i \hat{z}_{i, \uparrow} \hat{z}_{i, \downarrow}
    \end{split}
\end{equation}
where $a, a^{\dag}$ are the creation and annihilation operators respectively, $u$ is the repulsive interaction strength between spin-up and spin-down electrons at each site, and $\hat{z}$ is related to the number operator $\hat{n} = a^{\dag} a$ by $\hat{z} = (2\hat{n} - \mathbbm{1})$. The hopping strength is defined as $R_{i,j} = \tau$ if $i, j$ are nearest-neighbour lattice sites, allowing electrons to tunnel between adjacent sites, and $R_{i,j} = 0$ otherwise. We choose parameters that lie within the regime widely considered to be classically challenging for simulation \cite{campbellEarlyFaulttolerantSimulations2022, cadeStrategiesSolvingFermiHubbard2020}. As such, we set ${u}/{\tau}= 8$ and consider 2D square lattices of size $L \times L$ with $L \in \{4, 6, 8\}$. The total number of spin orbitals is given by $N = 2L^2$. 

To estimate the resource requirements for simulating the Fermi-Hubbard model using MMD, we employ the Fermionic swap network Trotter step algorithm to implement a second-order Trotterisation of the system, following Ref. \cite{kivlichanImprovedFaultTolerantQuantum2020}. We determine the number of rotations and corresponding angle of rotations required for each second-order Trotter step. Specifically, for $r$ Trotter steps, $N_{h} = 8Nr$ arbitrary rotations of angle $\theta_h = {\tau t}/{4r}$ are needed to simulate the hopping terms, while $N_{i} = {Nr}/{2}$ arbitrary rotations of angle $\theta_i = {ut}/{4r}$ are required to simulate the interaction terms. This makes use of a slight improvement upon Ref. \cite{kivlichanImprovedFaultTolerantQuantum2020} due to the shifted form of the interaction Hamiltonian \cite{campbellEarlyFaulttolerantSimulations2022}.

\begin{figure}[t]
    \centering
    \includegraphics[width=\linewidth]{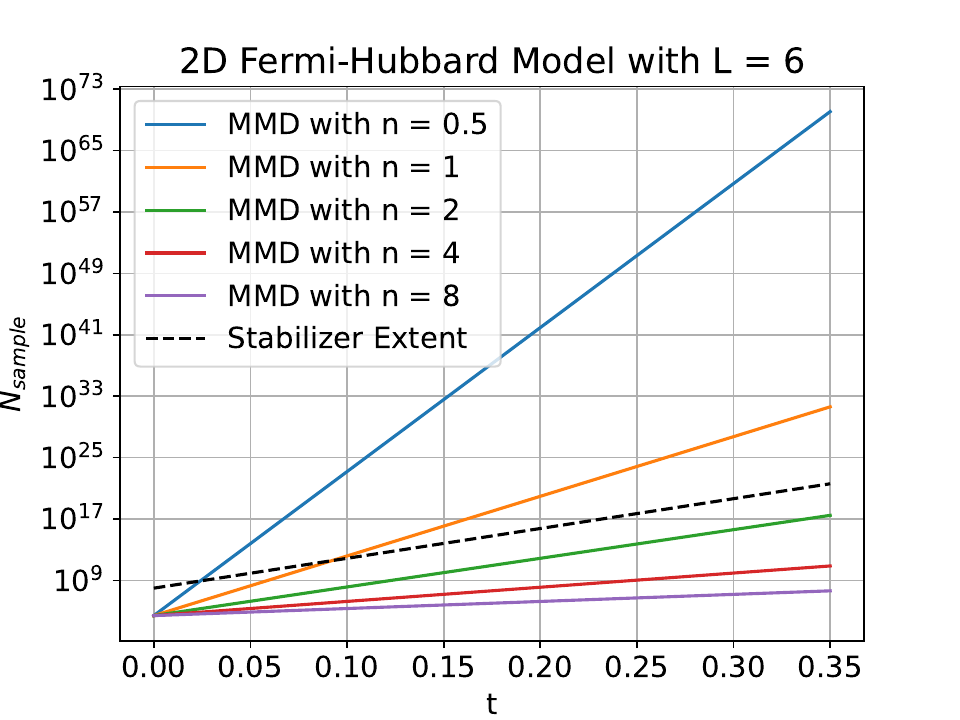}
    \caption{\label{fig:number of shots}Number of samples $(N_{\mathrm{sample}})$ required to perform MMD for the second-order Trotterised time evolution of the 2D Fermi-Hubbard model in the limit of large ($10^6$) Trotter steps.}
\end{figure}
\begin{figure*}[ht!]
    \centering
        \includegraphics[width=0.33\linewidth]{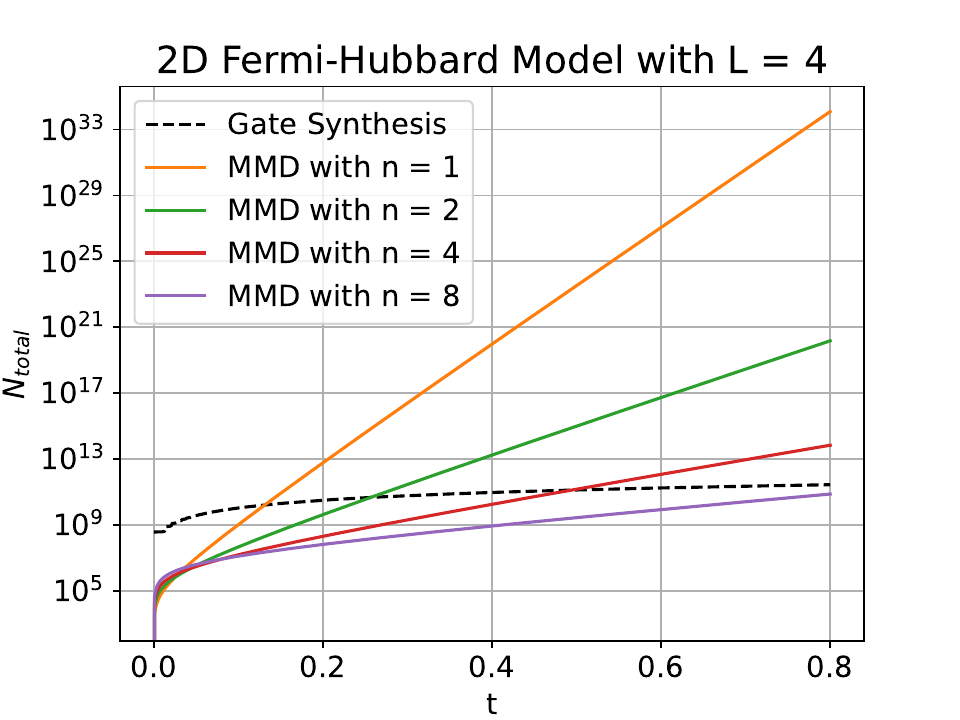}%
        \includegraphics[width=0.33\linewidth]{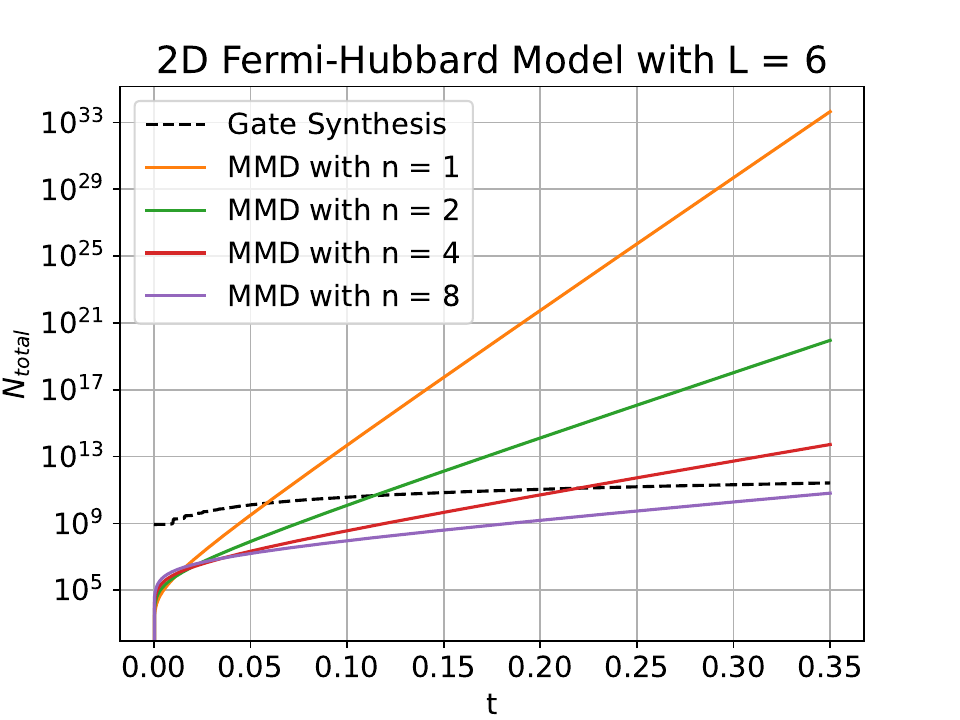}%
        \includegraphics[width=0.33\linewidth]{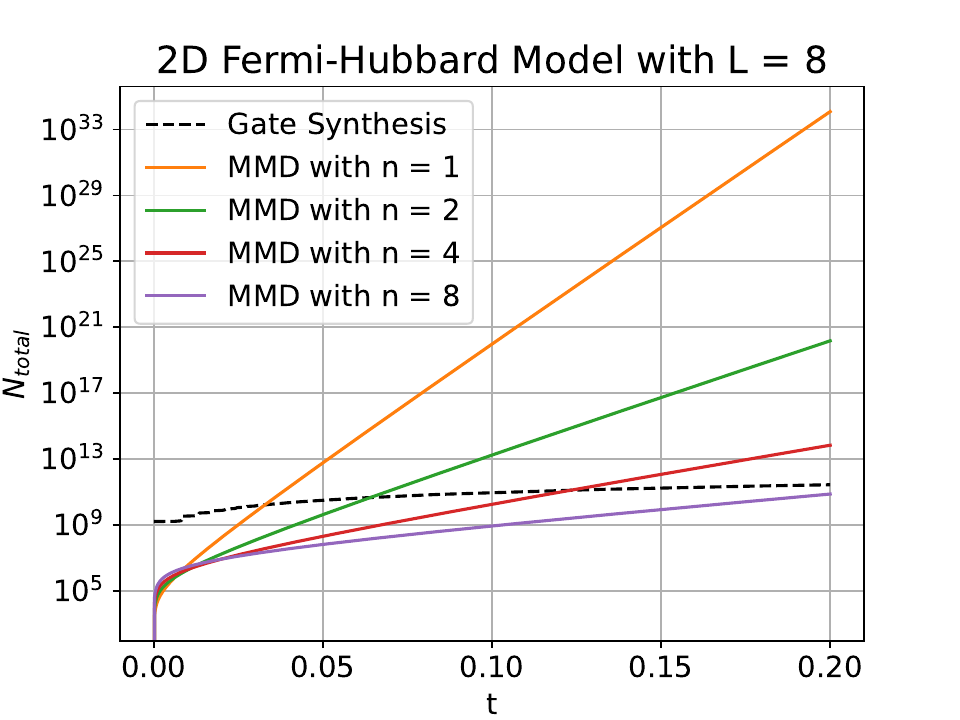}
    \caption{\label{fig:number of magic states over all shots}Number of magic states over all samples ($N_\text{total}$) to perform MMD for the second-order Trotterised time evolution of the 2D Fermi-Hubbard model compared to gate synthesis. Results are shown for an $L \times L$ lattice with $L \in \{4, 6, 8\}$.}
\end{figure*}
From this, we calculate the number of magic states required per sample as
\begin{equation}
    N_m = N_{h} E_{h} + N_{i} E_{i},
\end{equation}
where the number of magic states $E_{h(i)}$ (per sample) for each rotation is given by 
\begin{equation}
    E_{h(i)}= \bigg(2 - \frac{1}{n}\bigg) \frac{\big|x_n\big|}{\Lambda_\mathcal{G}(\mathcal{R}^{\theta_{h(i)}}_z)} \\
\end{equation}
in analogy to \cref{eqn:expected-magic-states}.  

The total number of samples ($N_{\mathrm{sample}}$) is calculated from Hoeffding's inequality to be 
\begin{equation}
    \begin{split}
        N_{\mathrm{sample}} = &\frac{2\ln ({2}/{\delta})}{\epsilon^2_{\mathrm{sample, EM}}}  \big[ (\Lambda_\mathcal{G}(\mathcal{R}^{\theta_{h}}_z))^{2N_{h}}
        (\Lambda_\mathcal{G}(\mathcal{R}^{\theta_{i}}_z))^{2N_{i}}\big].
    \end{split}
\end{equation}
Here we take the probability of error mitigation failing, $\delta$ to be $0.01$ and the error bound for failure mitigation $\epsilon_{\mathrm{sample, EM}}$ to be $0.02$. The total number of magic states over all samples then becomes 
\begin{equation}
    N_{\mathrm{total}} = N_m \times N_{\mathrm{sample}}.
\end{equation}
Considering the exponential scaling of the sampling overhead, we first evaluate the total number of samples for which our framework is practically feasible as a function of the evolution time $t$. We assume a dephasing noise of $p=0.1\%$ throughout the analysis in this section. In the limit of large Trotter steps for the MMD method, the number of samples for a lattice size of $L=6$ scales with $t$ as per \cref{fig:number of shots}. 

We see that a purely classical implementation of our MMD method (for which $n=0.5$) requires $10^{70}$ samples to simulate $t = 0.35$, whereas a quantum implementation of our method (so $n\geq 1$) requires from $10^{31}$ to $10^{7}$ samples as $n$ increases. We also overlay the upper-bounded runtime scaling of the stabilizer extent method with precision $\epsilon = 0.01$ in \cref{fig:number of shots}. 

We finally present a representative comparison of this framework with direct gate synthesis methods. Asymptotically optimal unitary synthesis is possible using the Ross-Selinger method \cite{rossOptimalAncillafreeClifford+T2016} with improvements possible by employing random compiling as shown in Ref. \cite{campbellShorterGateSequences2017}. Adding an ancillary qubit, mixed fallback techniques \cite{kliuchnikovShorterQuantum2023, yoshiokaErrorCrafting2025} obtain the best $T$-count known to date:
\begin{equation}
    T_{\mathrm{synth}} = 0.53 \log_2 \bigg( \frac{N_R}{\epsilon_{\mathrm{synth}}}\bigg) + 4.86,
\end{equation}
where $N_R = N_{h} + N_{i}$ is the total number of rotations. 

To ensure a fair comparison, we allow an error budget of $\epsilon_{\mathrm{synth}} + \epsilon_{\mathrm{trotter}} = 0.01$, a favourable sampling error $\epsilon_{\mathrm{sample, RS}} = 0.01$, and $N_{\mathrm{sample}}$ is a constant. Here, $\epsilon_{\mathrm{trotter}} = {W_{\mathrm{FS}} t^3}/{s^2}$ is the Trotter error for the Fermionic swap network algorithm, taking $W_{\mathrm{FS}}$ values from Ref. \cite{kivlichanImprovedFaultTolerantQuantum2020}. From this, we optimise the number of required Trotter steps $s$ that minimise $N_RT_{\mathrm{synth}}$ for gate synthesis. 

In \cref{fig:number of magic states over all shots}, we see that the MMD method provides a significant reduction in the magic state count for very small $t$, for all cases of $n$. Smaller lattice sizes see up to a few orders of magnitude saving and could provide an advantage for longer $t$ compared to direct gate synthesis. To achieve the same advantage as the lattice size and evolution time increases, MMD requires gates from higher levels (larger $n$) of the Clifford hierarchy as shown. In particular, for $L = 4, 6$ and $8$, MMD presents at least an order of magnitude resource saving for time evolution up to 0.8, 0.35 and 0.2 respectively. 

\section{\label{section:discussion}Discussion and Conclusions}
In this paper, we have presented our mitigated magic dilution framework for implementing small-angle single-qubit $Z$-rotations, making use of error mitigation to provide an advantage in terms of the sampling overhead when benchmarked against a classical approach. 

We have further identified a use-case of mitigated magic dilution for simulating time evolution of the Fermi-Hubbard Hamiltonian using Trotterisation. Through this, we have demonstrated a resource benefit of implementing mitigated magic dilution over conventional direct synthesis when taking into account the number of magic states and sampling overhead. Thus, our framework is suitable for early fault-tolerant quantum computers. Broadly speaking, our approach is more advantageous for algorithms containing a large number of very small angle rotations, such as Trotterised simulations. Our work hints towards a wide variety of directions for future work when larger angle rotations are encountered, including hybrid approaches with MMD used for small angles and synthesis used for larger angles, and hybrid approaches where the gate set $G$ used in MMD includes partially synthesised rotations.

A key assumption we have made in our MMD method is that all magic states are prepared with the same fidelity. Consequently, we observed optimal performance at some finite level of the Clifford hierarchy, fixed by the fidelity. However, recent literature indicates that this can be improved such that magic states of smaller rotations---and therefore gates higher in the Clifford hierarchy---can be prepared with a lower logical error rate \cite{choiFaultTolerantNonClifford2023, toshioPracticalQuantumAdvantage2024}. This is particularly relevant for this work as it suggests that the optimal performance will be at even higher Clifford hierarchy levels, leading to higher degrees of magic resource saving and hence speedup. Recent methods including magic state cultivation \cite{gidneyMagicStateCultivation2024} to construct high fidelity $|T\rangle$ magic states also complement the methods we present in this paper, however further research is needed to determine the feasibility and effectiveness of applying similar techniques to other magic states such as $|T^\frac{1}2{}\rangle$ for higher levels of the Clifford hierarchy.

It would be interesting to further analyse the resource requirements when compiling down to lattice surgery operations for particular algorithms.  Since our scheme uses repeated teleportation circuits (recall \cref{fig:generalised circuit}), these can be realised with logical $ZZ$ measurements without using Hadamard gates or patch rotations. In contrast, when using the gate synthesis approach, we require either Hadamard gates that take a long time to execute \cite{horsman2012surface,geher2024error}, or a fast data block structure that comes with an additional qubit overhead compared to compact layouts \cite{litinskiGameSurfaceCodes2019}.  

The Trotterised time evolution studied in this paper lends itself well into statistical phase estimation for future work. Furthermore, we have studied the effect of single-qubit dephasing noise channels, however for specific quantum hardware, it may be worthwhile to explore other noise models. Finally, further advancements in better logical magic states as discussed earlier will improve the advantages presented in this paper. 

We have compared our approach against two families of near-Clifford simulators.  Clearly, there are a range of other simulators to consider, including tensor network simulators that are suited to shallow local circuits on 2D arrays of qubits \cite{boixo2018characterizing}.  However, our techniques are directly applicable to simulating systems with long-range interactions or in higher-dimensional geometries, and in such settings tensor-network methods perform poorly.  Conversely, we envisage further improvements to mitigated magic dilution by extending to a dyadic decomposition approach similar to that used in some classical simulators \cite{seddonQuantifyingQuantumSpeedups2021}.    

\begin{acknowledgments}
This work was supported by the Engineering and Physical Sciences Research Council [grant numbers EP/S021582/1, EP/Y004140/1, EP/Y004310/1]. We thank Nick Blunt and Alex Gramolin for comments on the manuscript.
\end{acknowledgments}

\appendix
\onecolumngrid
\section{\label{app:A}Minimum Requirement of Three Channels}

In the general case, $A$ and $b$ in \cref{eqn:convex} can be written in matrix form as
\begin{equation}
    \label{eqn:MatA}
    A = 
    \begin{bmatrix}
        \cos^2{(\frac{\phi_1}{2})} & \cos^2{(\frac{\phi_2}{2})} & \cos^2{(\frac{\phi_3}{2})} & \cdots \\ \cos{(\frac{\phi_1}{2})}\sin{(\frac{\phi_1}{2})} & \cos{(\frac{\phi_2}{2})}\sin{(\frac{\phi_2}{2})} & \cos{(\frac{\phi_3}{2})}\sin{(\frac{\phi_3}{2})} & \cdots \\ \sin^2{(\frac{\phi_1}{2})} & \sin^2{(\frac{\phi_2}{2})} & \sin^2{(\frac{\phi_3}{2})} & \cdots
    \end{bmatrix},
\end{equation}
and
\begin{equation}
    \label{eqn:Vecb}
    b = 
    \begin{bmatrix}
        \cos^2{(\frac{\theta}{2})} \\ \cos{(\frac{\theta}{2})}\sin{(\frac{\theta}{2})} \\ \sin^2{(\frac{\theta}{2})}.
    \end{bmatrix}.
\end{equation}
Specifically for the two channel case, we have
\begin{equation}
    \begin{bmatrix}
        \cos^2{(\frac{\phi_1}{2})} & \cos^2{(\frac{\phi_2}{2})} \\ \cos{(\frac{\phi_1}{2})}\sin{(\frac{\phi_1}{2})} & \cos{(\frac{\phi_2}{2})}\sin{(\frac{\phi_2}{2})} \\ \sin^2{(\frac{\phi_1}{2})} & \sin^2{(\frac{\phi_2}{2})}
    \end{bmatrix}
    \begin{bmatrix}
        x_1 \\ x_2 
    \end{bmatrix}
    = \begin{bmatrix}
        \cos^2{(\frac{\theta}{2})} \\ \cos{(\frac{\theta}{2})}\sin{(\frac{\theta}{2})} \\ \sin^2{(\frac{\theta}{2})},
    \end{bmatrix}
\end{equation}
from which we write three simultaneous equations:
\begin{subequations}
    \begin{align}
    x_1  \cos^2{\bigg(\frac{\phi_1}{2}\bigg)} + x_2 \cos^2{\bigg(\frac{\phi_2}{2}\bigg)} =& \cos^2{\bigg(\frac{\theta}{2}\bigg)},  \label{eqn:A4a} \\
    x_1 \cos{\bigg(\frac{\phi_1}{2}\bigg)}\sin{\bigg(\frac{\phi_1}{2}\bigg)} + x_2 \cos{\bigg(\frac{\phi_2}{2}\bigg)}\sin{\bigg(\frac{\phi_2}{2}\bigg)} =& \cos{\bigg(\frac{\theta}{2}\bigg)}\sin{\bigg(\frac{\theta}{2}\bigg)}, \label{eqn:A4b} \\
    x_1  \sin^2{\bigg(\frac{\phi_1}{2}\bigg)} + x_2 \sin^2{\bigg(\frac{\phi_2}{2}\bigg)} =& \sin^2{\bigg(\frac{\theta}{2}\bigg)}. \label{eqn:A4c} 
    \end{align}
    \label{eqn:system of two channels}
\end{subequations}

Upon solving \cref{eqn:A4a} and \cref{eqn:A4c}, we obtain
\begin{equation}
    x_1 + x_2 = 1,
\end{equation}
which arises due to the unitarity of the channels. From this, we can find two solutions for $x_1$ by substituting this into \cref{eqn:A4a} and \cref{eqn:A4b} respectively, resulting in
\begin{equation}
    \begin{split}
        x_1 =& \frac{\cos((\theta+\phi_1)/2) \sin((\theta + \phi_2)/2)}{\cos((\phi_1 + \phi_2)/2)\sin((\phi_1+\phi_2)/2)} \\
        =& \frac{\sin((\theta+\phi_1)/2) \sin((\theta + \phi_2)/2)}{\sin((\phi_1 + \phi_2)/2)\sin((\phi_1+\phi_2)/2)}.
    \end{split}
\end{equation}
Therefore all three equations in this overdetermined system are only satisfied when 
\begin{equation}
    \phi_1 = \theta + 2k\pi \Rightarrow x_1 = 1, x_2 = 0
\end{equation}
or 
\begin{equation}
    \phi_2 = \theta + 2k\pi \Rightarrow x_1 = 0, x_2 = 1,
\end{equation}
where $k \in \mathbb{Z}$, i.e., the trivial case of only a single channel in the decomposition. So, for multiple channels in the decomposition, at least three channels are required. 

\section{LCC Decomposition for $\{\mathcal{I}, {\mathcal{T}}^\frac{1}{n}, \mathcal{Z}\}$ Channels} 

In the main text, we showed that the optimal decompositions found using convex optimisation were of the form $\{\mathcal{I}, {\mathcal{T}}^\frac{1}{n}, \mathcal{Z}\}$ for both ideal and noisy $\mathcal{T}^\frac{1}{n}$ channels. In the following sections, we derive the $l_1$-norm and $\gamma$ corresponding to these optimal decompositions.

\subsection{Ideal Case}
\label{app:B1}

A $\mathcal{T}^{\frac{1}{n}}$ channel can be written as a $\mathcal{Z}$-rotation channel with rotation angle $\phi$,
\begin{equation}
    \begin{split}
        \mathcal{R}_z^{\phi}(\rho) =&\ e^{-i\frac{\phi}{2} Z} \rho e^{i\frac{\phi}{2} Z} \\
        =&\ \cos^2{\bigg(\frac{\phi}{2}\bigg)} \rho +\sin^2{\bigg(\frac{\phi}{2}\bigg)} Z\rho Z + i \cos{\bigg(\frac{\phi}{2}\bigg)}\sin{\bigg(\frac{\phi}{2}\bigg)} [\rho Z - Z\rho],
    \end{split}
\end{equation}
 where $\phi = \frac{\pi}{4n}$. Recall that $n = 2^{i-1}$ where $i \in \mathbb{Z_{\geq \text{0}}}$ such that $i = 0$ is the $\mathcal{S}$ channel, $i = 1$ is the $\mathcal{T}$ channel, etc. 

For an LCC decomposition in terms of $\{\mathcal{I}, \mathcal{R}_z^{\phi}, \mathcal{Z}\}$ channels, we obtain
\begin{equation}
    \begin{split}
         \mathcal{R}^{\theta}_z(\rho) &= x_0\mathcal{I}(\rho) + x_1\mathcal{R}^{\phi}_z(\rho) + x_2\mathcal{Z}(\rho) \\
         &= x_0\rho + x_1 \bigg(\cos^2{\bigg(\frac{\phi}{2}\bigg)} \rho + i \cos{\bigg(\frac{\phi}{2}\bigg)}\sin{\bigg(\frac{\phi}{2}\bigg)} [\rho Z - Z\rho] + \sin^2{\bigg(\frac{\phi}{2}\bigg)} Z\rho Z \bigg) + x_2 Z\rho Z \nonumber \\
         &= \bigg(x_0 + \cos^2{\bigg(\frac{\phi}{2}\bigg)} x_1 \bigg) \rho + i\cos{\bigg(\frac{\phi}{2}\bigg)}\sin{\bigg(\frac{\phi}{2}\bigg)} x_1 [\rho Z - Z\rho] +  \bigg(\sin^2{\bigg(\frac{\phi}{2}\bigg)} x_1 + x_2 \bigg) Z\rho Z \nonumber \\
         &= \cos^2{\bigg(\frac{\theta}{2}\bigg)} \rho + i \cos{\bigg(\frac{\theta}{2}\bigg)}\sin{\bigg(\frac{\theta}{2}\bigg)} [\rho Z - Z\rho] + \sin^2{\bigg(\frac{\theta}{2}\bigg)} Z\rho Z.
    \end{split}
\end{equation}
The coefficients $x_i$ can be determined by solving three simultaneous equations: 
\begin{subequations}
    \begin{align}
        x_0 + \cos^2{\bigg(\frac{\phi}{2}\bigg)} x_1 &= \cos^2{\bigg(\frac{\theta}{2}\bigg)}, \\
        \cos{\bigg(\frac{\phi}{2}\bigg)}\sin{\bigg(\frac{\phi}{2}\bigg)} x_1 &= \cos{\bigg(\frac{\theta}{2}\bigg)}\sin{\bigg(\frac{\theta}{2}\bigg)}, \\
        \sin^2{\bigg(\frac{\phi}{2}\bigg)} x_1 + x_2&= \sin^2{\bigg(\frac{\theta}{2}\bigg)},
    \end{align}
\end{subequations}
with solutions given by
\begin{subequations}
    \begin{align}
        x_0 &= \cos^2{\bigg(\frac{\theta}{2}\bigg)} - \bigg(\frac{\sin(\theta)}{\sin(\phi)} \bigg)\cos^2{\bigg(\frac{\phi}{2}\bigg)},  \\
        x_1 &= \frac{\sin(\theta)}{\sin(\phi)}, \\
        x_2 &= \sin^2{\bigg(\frac{\theta}{2}\bigg)} - \bigg(\frac{\sin(\theta)}{\sin(\phi)} \bigg)\sin^2{\bigg(\frac{\phi}{2}\bigg)}.
    \end{align}
\end{subequations}
The $l_1$-norm of this decomposition for target rotation angles $0 \leq \theta \leq \phi$ is 
\begin{equation}
    \begin{split}
        \Lambda_\mathcal{G}(\mathcal{R}^\theta_z) &= |x_0| + |x_1| + |x_2| \\
        &= \bigg| \cos^2{\bigg(\frac{\theta}{2}\bigg)} - \bigg(\frac{\sin(\theta)}{\sin(\phi)} \bigg)\cos^2{\bigg(\frac{\phi}{2}\bigg)}\bigg| 
        + \bigg| \frac{\sin(\theta)}{\sin(\phi)} \bigg| + \bigg| \sin^2{\bigg(\frac{\theta}{2}\bigg)} - \bigg(\frac{\sin(\theta)}{\sin(\phi)} \bigg)\sin^2{\bigg(\frac{\phi}{2}\bigg)} \bigg| \\
        &= \cos^2{\bigg(\frac{\theta}{2}\bigg)} - \bigg(\frac{\sin(\theta)}{\sin(\phi)} \bigg)\cos^2{\bigg(\frac{\phi}{2}\bigg)} + \frac{\sin(\theta)}{\sin(\phi)}  - \sin^2{\bigg(\frac{\theta}{2}\bigg)} + \bigg(\frac{\sin(\theta)}{\sin(\phi)} \bigg)\sin^2{\bigg(\frac{\phi}{2}\bigg)}  \\
        &= \cos{(\theta)} + \frac{\sin{(\theta)}}{\sin{(\phi)}} \bigg(1 - \cos{(\phi)}\bigg).
    \end{split}
\end{equation}
noting that $x_2 \leq 0$ while $x_0, x_1 \geq 0$ for this range of $\theta$.

For small rotation angles $\theta$, we approximate
\begin{equation}
    \begin{split}
        \Lambda_\mathcal{G}(\mathcal{R}^\theta_z) &= 
        \cos{(\theta)} + \sin{(\theta)} \bigg(\frac{1 - \cos{(\phi)}}{\sin{(\phi)}} \bigg) \\
        &\approx 1 - \frac{\theta^2}{2} + \theta \bigg(\frac{1 - \cos{(\phi)}}{\sin{(\phi)}} \bigg) \\
        &\approx 1 + \theta \bigg(\frac{1 - \cos{(\phi)}}{\sin{(\phi)}} \bigg).
    \end{split}
\end{equation}
Thus $\gamma$ in \cref{eqn:gamma defn} is obtained as follows
\begin{equation}
    \begin{split}
        \gamma(\theta) &= \frac{\ln(\Lambda_\mathcal{C}(\mathcal{R}_z^\theta))}{\ln(\Lambda_\mathcal{G}(\mathcal{R}_z^\theta))} \\
        &= \frac{\ln{(1 + \theta)}}{\ln{\bigg(1 + \theta \big(\frac{1 - \cos{(\phi)}}{\sin{(\phi)}} \big)}\bigg)} \\
        &\approx\frac{\theta}{\big(\frac{1 - \cos{(\phi)}}{\sin{(\phi)}}\big)\theta} \\
        &= \frac{\sin{(\phi)}}{1 - \cos{(\phi)}}  = \cot\bigg({\frac{\phi}{2}}\bigg).
    \end{split}
\end{equation}

\subsection{Dephased $\mathcal{T}^\frac{1}{n}$ Case}
\label{app:B2}

We consider a dephasing noise channel as given by \cref{eqn:dephased U(p)}. In the main text, we define an effective dephasing noise $p_\text{eff}$, which we use in the following derivation. 

The dephasing noise channel can applied to $\mathcal{R}_z^{\phi}$ as follows:
\begin{equation}
\begin{split}
    \varepsilon(\mathcal{R}_z^{\phi}(\rho)) =& (1 - p_\text{eff})(\mathcal{R}_z^{\phi}(\rho)) + p_\text{eff}Z(\mathcal{R}_z^{\phi}(\rho))Z \\
    =& \bigg[ \cos^2{\bigg(\frac{\phi}{2}\bigg)} - p_\text{eff} \cos{(\phi)} \bigg] \rho + (1-2p_\text{eff})\bigg[i \cos{\bigg(\frac{\phi}{2}\bigg)}\sin{\bigg(\frac{\phi}{2}\bigg)} \bigg] [\rho Z - Z \rho] \\
    &+ \bigg[ \sin^2{\bigg(\frac{\phi}{2}\bigg)} + p_\text{eff} \cos{(\phi)} \bigg] Z \rho Z .
\end{split}
\end{equation}
The LCC decomposition of the $\mathcal{R}_z^{\theta}(\rho)$ channel in terms of $\{\mathcal{I}, \varepsilon(\mathcal{R}_z^{\phi}), \mathcal{Z}\}$ is given by 
\begin{subequations}
\begin{align}
     \mathcal{R}_z^{\theta}(\rho) =& \ x_0\mathcal{I}(\rho) + x_1\varepsilon(\mathcal{R}_z^{\phi}(\rho)) + x_2\mathcal{Z}(\rho) \\
     =& \ x_0\rho + x_1 \Bigg[\bigg[ \cos^2{\bigg(\frac{\phi}{2}\bigg)} - p_\text{eff} \cos{(\phi)} \bigg] \rho + (1-2p_\text{eff})\bigg[i \cos{\bigg(\frac{\phi}{2}\bigg)}\sin{\bigg(\frac{\phi}{2}\bigg)} \bigg] [\rho Z - Z \rho] \nonumber \\ &+ \bigg[ \sin^2{\bigg(\frac{\phi}{2}\bigg)} + p_\text{eff}\cos{(\phi)} \bigg] Z \rho Z \Bigg] + x_2 Z\rho Z \nonumber \\
     =& \bigg(x_0 + \bigg[ \cos^2{\bigg(\frac{\phi}{2}\bigg)} - p_\text{eff} \cos{(\phi)} \bigg] x_1 \bigg) \rho + i(1-2p_\text{eff})\cos{\bigg(\frac{\phi}{2}\bigg)}\sin{\bigg(\frac{\phi}{2}\bigg)} x_1 [\rho Z - Z\rho] \nonumber \\ &+\bigg( \bigg[ \sin^2{\bigg(\frac{\phi}{2}\bigg)} + p_\text{eff} \cos{(\phi)}\bigg]x_1 + x_2 \bigg) Z\rho Z \nonumber \\
     =& \cos^2{\bigg(\frac{\theta}{2}\bigg)} \rho + i \cos{\bigg(\frac{\theta}{2}\bigg)}\sin{\bigg(\frac{\theta}{2}\bigg)} [\rho Z - Z\rho] + \sin^2{\bigg(\frac{\theta}{2}\bigg)} Z\rho Z
\end{align}
\end{subequations}
The coefficients $x_i$ can be determined by solving three simultaneous equations: 
\begin{subequations}
\begin{align}
    x_0 + \bigg[ \cos^2{\bigg(\frac{\phi}{2}\bigg)} - p_\text{eff} \cos{(\phi)} \bigg] x_1 &= \cos^2{\bigg(\frac{\theta}{2}\bigg)},\\
    (1-2p_\text{eff})\cos{\bigg(\frac{\phi}{2}\bigg)}\sin{\bigg(\frac{\phi}{2}\bigg)} x_1 &= \cos{\bigg(\frac{\theta}{2}\bigg)}\sin{\bigg(\frac{\theta}{2}\bigg)}, \\
    \bigg[ \sin^2{\bigg(\frac{\phi}{2}\bigg)} + p_\text{eff} \cos{(\phi)}\bigg]x_1 + x_2&= \sin^2{\bigg(\frac{\theta}{2}\bigg)}.
\end{align}
\end{subequations}
with solutions given by 
\begin{subequations}
\begin{align}
    x_0 &= \cos^2{\bigg(\frac{\theta}{2}\bigg)} - \bigg[ \cos^2{\bigg(\frac{\phi}{2}\bigg)} - p_\text{eff} \cos{(\phi)} \bigg]\frac{\sin(\theta)}{(1-2p_\text{eff})\sin(\phi)}\\
    x_1 &= \frac{\sin(\theta)}{(1-2p_\text{eff})\sin(\phi)} \\
    x_2 &= \sin^2{\bigg(\frac{\theta}{2}\bigg)} - \bigg[ \sin^2{\bigg(\frac{\phi}{2}\bigg)} + p_\text{eff} \cos{(\phi)}\bigg]\frac{\sin(\theta)}{(1-2p_\text{eff})\sin(\phi)}.
\end{align}
\end{subequations}
The $l_1$-norm of this decomposition is
\begin{equation}
\begin{split}
    \Lambda_\mathcal{G}(\mathcal{R}^\theta_z) =& \ |x_0| + |x_1| + |x_2| \\
    =& \ \bigg| \cos^2{\bigg(\frac{\theta}{2}\bigg)} - \bigg[ \cos^2{\bigg(\frac{\phi}{2}\bigg)} - p_\text{eff} \cos{(\phi)} \bigg]\frac{\sin(\theta)}{(1-2p_\text{eff})\sin(\phi)}\bigg| 
    + \bigg| \frac{\sin(\theta)}{(1-2p_\text{eff})\sin(\phi)}  \bigg| \\
    &+ \bigg| \sin^2{\bigg(\frac{\theta}{2}\bigg)} - \bigg[ \sin^2{\bigg(\frac{\phi}{2}\bigg)} + p_\text{eff}\cos{(\phi)}\bigg]\frac{\sin(\theta)}{(1-2p_\text{eff})\sin(\phi)}\bigg| \\
    =& \ \cos^2{\bigg(\frac{\theta}{2}\bigg)} - \bigg[ \cos^2{\bigg(\frac{\phi}{2}\bigg)} - p_\text{eff} \cos{(\phi)} \bigg]\frac{\sin(\theta)}{(1-2p_\text{eff})\sin(\phi)} 
    + \frac{\sin(\theta)}{(1-2p_\text{eff})\sin(\phi)} \\
    &- \sin^2{\bigg(\frac{\theta}{2}\bigg)} + \bigg[ \sin^2{\bigg(\frac{\phi}{2}\bigg)} + p_\text{eff}\cos{(\phi)}\bigg]\frac{\sin(\theta)}{(1-2p_\text{eff})\sin(\phi)} \\
    =& \ \cos{(\theta)} - \frac{\sin(\theta)}{(1-2p_\text{eff})\sin(\phi)} \bigg((1-2p_\text{eff})\cos{(\phi)} - 1 \bigg) \\
    =& \cos{(\theta)} + \sin(\theta) \bigg(\frac{1}{(1-2p_\text{eff})}\csc{(\phi)} - \cot{(\phi)} \bigg). 
\end{split}
\end{equation}
for target rotation angles $0 \leq \theta \ll \phi$. 
For small $\theta$, we can approximate
\begin{equation}
    \begin{split}
        \Lambda_\mathcal{G}(\mathcal{R}^\theta_z) 
        &= \cos{(\theta)} + \sin(\theta) \bigg(\frac{1}{(1-2p_\text{eff})}\csc{(\phi)} - \cot{(\phi)} \bigg) \\
        &\approx 1 - \frac{\theta^2}{2} + \theta \bigg( \frac{1}{(1-2p_\text{eff})}\csc{(\phi)} - \cot{(\phi)} \bigg) \\
        &\approx 1 + \theta \bigg( \frac{1}{(1-2p_\text{eff})}\csc{(\phi)} - \cot{(\phi)} \bigg).
    \end{split}
\end{equation}
Therefore, $\gamma$ becomes
\begin{equation}
    \begin{split}
        \gamma(\theta) &= \frac{\ln(\Lambda_\mathcal{C}(\mathcal{R}_z^\theta))}{\ln(\Lambda_\mathcal{G}(\mathcal{R}_z^\theta))} \\
        &= \frac{\ln{(1+\theta )}}{\ln{\bigg(1 + \theta \bigg( \frac{1}{(1-2p_\text{eff})}\csc{(\phi)} - \cot{(\phi)} \bigg)}\bigg)} \\
        &\approx \frac{\theta}{\theta \bigg( \frac{1}{(1-2p_\text{eff})}\csc{(\phi)} - \cot{(\phi)}\bigg)} \\
        &= \frac{1}{\frac{\csc{(\phi)}}{(1-2p_\text{eff})} - \cot{(\phi)}}.
    \end{split}
\end{equation}

\section{Noise in Generalised Gate Teleportation Circuit}
\label{app:C}
We can show that the generalised gate teleportation circuit in \cref{fig:generalised circuit} results in a rotation that differs from the target rotation $\theta$ by either dephasing noise, or coherent errors. We note that coherent errors can be handled through calibration, and accurate calibration is an underlying assumption required for the estimation of the level of dephasing noise present. 

We proceed by considering a single step of generalised teleportation with noisy states as shown in \cref{fig:noisy_teleportation}.
\begin{figure}[h]
    \centering
    \begin{quantikz}
    \rho \ & \ctrl{1} & \gate[style = dashed]{\sqrt[n/2]{T}} & \ \mathcal{E}(\rho) \approx \sqrt[n]{T} \rho \sqrt[n]{T}^\dagger  \\
    \sigma_n \approx \vert \sqrt[n]{T}\rangle \langle \sqrt[n]{T} \vert & \targ{}  & \meter{} \wire[u][1]{c}
    \end{quantikz}
    \caption{\label{fig:noisy_teleportation} A single step of generalised teleportation when the magic state used is an arbitrary mixed state $\sigma_n$. The outcome of the measurement is denoted $m$ such that the correction is applied if $m=1$ (eigenvalue is $-1$).}
\end{figure}

The state $\sigma_n$ will have an eigenvalue decomposition
\begin{align}
  \sigma_n = \vert \phi_0 \rangle \langle \phi_0 \vert +  \vert \phi_1 \rangle \langle \phi_1 \vert
\end{align}
where $\vert \phi_j \rangle$ are not individually normalised, but $\langle \phi_0 \vert  \phi_0 \rangle + \langle \phi_1 \vert  \phi_1 \rangle$ = 1. It is a common assumption to take $\vert \phi_0 \rangle = \sqrt{1-p} \vert \sqrt[n]{T}^\dagger \rangle$ and $\vert \phi_1 \rangle$ as its orthogonal partner.  However, the purpose of this appendix is to allow for a fully general $\sigma_n$.

After measurement outcome $m$, $\rho$ gets mapped to 
\begin{equation}
    \mathcal{E}_{m}(\rho) = \kappa_{0,m} \rho \kappa_{0,m}^\dagger  + \kappa_{1,m} \rho \kappa_{1,m}^\dagger ,
\end{equation}
where $\kappa_{j,m}$ is the Kraus operator corresponding to teleportation using a pure state $\vert \phi_j \rangle$ after measurement outcome $m$.  Denoting $\vert \phi_j \rangle = A_j  \vert 0 \rangle + B_j \vert 1 \rangle$, calculating the action of the teleportation circuit one finds
\begin{subequations}
\begin{align}
    \kappa_{0,0} & =  A_0 \vert 0 \rangle \langle 0 \vert  + B_0  \vert 1 \rangle \langle 1 \vert   = \left( \frac{A_0 + B_0}{2} \right) I  + \left( \frac{A_0 - B_0}{2} \right) Z  , \\ 
    \kappa_{0,1} & =  A_0 \vert 1 \rangle \langle 1 \vert + B_0  \vert 0 \rangle \langle 0 \vert     = \left( \frac{A_0 + B_0}{2} \right) I  + \left( \frac{-A_0 + B_0}{2} \right) Z  , \\  
     \kappa_{1,0} & =   A_1 \vert 0 \rangle \langle 0 \vert+ B_1 \vert 1 \rangle \langle 1 \vert   = \left( \frac{A_1 + B_1}{2} \right) I  + \left( \frac{A_1 - B_1}{2} \right) Z  , \\ 
    \kappa_{1,1} & =  A_1 \vert 1 \rangle \langle 1 \vert + B_1  \vert 0 \rangle \langle 0 \vert     = \left( \frac{A_1 + B_1}{2} \right) I  + \left( \frac{-A_1 + B_1}{2} \right) Z  
\end{align}
\end{subequations}
and so every relevant Kraus operator is diagonal in the $Z$-basis.  Therefore, expanding out the channel we find that
\begin{align}
    \mathcal{E}_{m}(\rho) = E_{0,0}^{m} \rho + E_{0,1}^{m} \rho Z + E_{1,0}^{m} Z \rho  + E_{1,1}^m Z\rho Z.
\end{align}
where $E_{k,\ell}^m$ are numbers depending on $A_0, A_1, B_0, A_1$ that in turn depend on $\sigma_n$.

We are not quite done because in the case where $m=1$, we need to apply a correction (i.e. using a magic state from one level lower in the Clifford hierarchy $\sigma_{n/2}$). Nonetheless, this correction process will also perform diagonal Kraus operators, and since the group of diagonal operators is closed in multiplication, we conclude that generalised teleportation with result in some channel $\mathcal{E}$ that is a sum of diagonal Kraus operators.  Expanding these out in the Pauli basis, we will again have an expression of the form
\begin{align} \label{Eform}
    \mathcal{E}(\rho) = E_{0,0} \rho + E_{0,1} \rho Z + E_{1,0} Z \rho  + E_{1,1} Z\rho Z,
\end{align}
where $E_{i,j}$ depend on the density matrices of all magic states $\{ \sigma_{n}, \sigma_{n/2}, \sigma_{n/4}, \ldots, \sigma_{1} \}$ in the generalised teleportation circuit.  This means a large number of parameters are involved, but fortunately we can proceed with our proof using only the form of \cref{Eform} and the following observation: once all corrections are completed (we mix over all measurement outcomes) the full channel $\mathcal{E}$ must be CPTP. Note that, in contrast, each component $\mathcal{E}_m$ will be completely positive (CP) but not necessarily trace preserving (TP). Therefore, under the Choi–Jamiołkowski isomorphism, the Choi state for this channel must be a physical density matrix (Hermitian, positive semi-definite and trace normalized to unity).  Furthermore, due to the form of \cref{Eform}, the Choi state has the form  
\begin{align} \label{Choi}
 \Phi_{\mathcal{E}} = (\mathcal{E} \otimes \mathcal{I})( \vert \Phi \rangle  \langle \Phi \vert ) =  \left( \begin{array}{cccc}
a_{0,0} & 0 & 0 & a_{0,1} \\
0 & 0 & 0 & 0 \\
0 & 0 & 0 & 0 \\
a_{1,0} & 0 & 0 & a_{1,1} \\
 \end{array}\right). 
\end{align}
where for instance $a_{1,1}=(E_{0,0}-E_{0,1}-E_{1,0}+E_{1,1})/2$. 
In particular, the Choi state has at most 2 non-zero eigenvalues, and by the CPTP property we can denote these as $p$ and $1-p$ (with $1\leq p \leq 0$), so they can be interpreted as probabilities. Therefore,  
\begin{align}
   \Phi_{\mathcal{E}} = p \vert K_0 \rangle \langle K_0 \vert  + (1-p) \vert K_1 \rangle \langle K_1 \vert  
\end{align}
where $\vert K_j \rangle$ are a pair of orthogonal pure states supported on the non-trivial $2 \times 2$ submatrix of \cref{Choi}. We may now reverse the Choi–Jamiołkowski isomorphism, so that 
\begin{align}
   \mathcal{E}(\rho) = p  K_0 \rho K_0^\dagger  + (1-p) K_1 \rho K_1^\dagger
\end{align}
where $K_j$ is the single Kraus operator isomorphic to the state $\vert K_j \rangle \langle K_j \vert$. Since the state representation is restricted to a specific submatrix, we can conclude that $K_j$ are unitary operators diagonal in the $Z$ basis and we are free to choose the global phase. Therefore, there exists angles $\varphi_j$ such that $K_{j} = R_{z}(\varphi_j)$. Finally, since $\vert K_j \rangle$ are orthogonal to each other, we can conclude that $\mathrm{Tr}[K_{1}^\dagger K_{0} ]=0$.  This orthogonality entails that $\varphi_1$ is such that $K_1 = K_0 Z$ upto a global phase. This brings us to the final form
\begin{align}
  \mathcal{E}(\rho) = p \mathcal{R}_z^{\varphi_0}(\rho) + (1-p )Z \mathcal{R}_z^{\varphi_0}(\rho) Z
\end{align}
which has the claimed form of dephasing noise and potentially a coherent error which deviates from the ideal angle $\phi$ by a phase of $\delta = \varphi_0-\phi$.  

We have assumed throughout that physical noise is well characterized. The impact of changes in the magic states $\sigma_n$ affects the form of the resulting channel $\mathcal{E}$.  If there is a undesired coherent error, we can adjust the angle of the prepared magic state $\sigma_n$ to eliminate this, thereby leaving only dephasing error as assumed throughout the main text.  
\section{Supplementary Data}
\label{app:D}

\begin{table}[H]
\caption{\label{}$\ln(\Lambda_{G}(\mathcal{R}_z^{\theta}))$ as a function of target rotation angle $\theta$ for different optimal decompositions of the form $\{\mathcal{I}, \varepsilon(\mathcal{T}^\frac{1}n{}), \mathcal{Z}\}$ including the optimal Clifford decomposition of $\{\mathcal{I}, \mathcal{S}, \mathcal{Z}\}$. The values presented here correspond to the relative values used to calculate $\gamma$ in \cref{table:1}.}

\centering
\begin{tabular}{|c|c|c|c|c|c|c|}
\hline
$n$ & $\phi$ & $\mathcal{R}_z^{\phi}$ & \multicolumn{4}{|c|}{$\ln(\Lambda_\mathcal{G}(\mathcal{R}^\theta_z))$ (Small $\theta$)} \\
\hline
& & & $p = 0.01\%$ & $p = 0.1\%$ & $p = 0.5\%$ & $p = 1.0\%$ \\
\hline
0 & $\frac{\pi}{2}$ & $\mathcal{S}$ & 1.00E-07 & 1.00E-07 & 1.00E-07 & 1.00E-07  \\
1 & $\frac{\pi}{4}$ & $\mathcal{T}$ & 4.14E-08 & 4.17E-08 & 4.28E-08 & 4.43E-08  \\
2 & $\frac{\pi}{8}$ & $\mathcal{\sqrt{T}}$ & 2.00E-08 & 2.07E-08 & 2.39E-08 & 2.40E-08 \\
4 & $\frac{\pi}{16}$ & $\mathcal{\sqrt[4]{T}}$ & 1.00E-08 & 1.16E-08 & 1.90E-08 & 2.84E-08 \\
8 & $\frac{\pi}{32}$ & $\mathcal{\sqrt[8]{T}}$ & 5.30E-09 & 8.75E-09 & 2.44E-08 & 4.47E-09\\
\hline
\end{tabular}
\end{table}

\twocolumngrid
\bibliography{bibliography}

\begin{thebibliography}{59}%
\makeatletter
\providecommand \@ifxundefined [1]{%
 \@ifx{#1\undefined}
}%
\providecommand \@ifnum [1]{%
 \ifnum #1\expandafter \@firstoftwo
 \else \expandafter \@secondoftwo
 \fi
}%
\providecommand \@ifx [1]{%
 \ifx #1\expandafter \@firstoftwo
 \else \expandafter \@secondoftwo
 \fi
}%
\providecommand \natexlab [1]{#1}%
\providecommand \enquote  [1]{``#1''}%
\providecommand \bibnamefont  [1]{#1}%
\providecommand \bibfnamefont [1]{#1}%
\providecommand \citenamefont [1]{#1}%
\providecommand \href@noop [0]{\@secondoftwo}%
\providecommand \href [0]{\begingroup \@sanitize@url \@href}%
\providecommand \@href[1]{\@@startlink{#1}\@@href}%
\providecommand \@@href[1]{\endgroup#1\@@endlink}%
\providecommand \@sanitize@url [0]{\catcode `\\12\catcode `\$12\catcode `\&12\catcode `\#12\catcode `\^12\catcode `\_12\catcode `\%12\relax}%
\providecommand \@@startlink[1]{}%
\providecommand \@@endlink[0]{}%
\providecommand \url  [0]{\begingroup\@sanitize@url \@url }%
\providecommand \@url [1]{\endgroup\@href {#1}{\urlprefix }}%
\providecommand \urlprefix  [0]{URL }%
\providecommand \Eprint [0]{\href }%
\providecommand \doibase [0]{https://doi.org/}%
\providecommand \selectlanguage [0]{\@gobble}%
\providecommand \bibinfo  [0]{\@secondoftwo}%
\providecommand \bibfield  [0]{\@secondoftwo}%
\providecommand \translation [1]{[#1]}%
\providecommand \BibitemOpen [0]{}%
\providecommand \bibitemStop [0]{}%
\providecommand \bibitemNoStop [0]{.\EOS\space}%
\providecommand \EOS [0]{\spacefactor3000\relax}%
\providecommand \BibitemShut  [1]{\csname bibitem#1\endcsname}%
\let\auto@bib@innerbib\@empty
\bibitem [{\citenamefont {{Google Quantum AI and Collaborators}}(2024)}]{acharyaQuantumErrorCorrection2024}%
  \BibitemOpen
  \bibfield  {author} {\bibinfo {author} {\bibnamefont {{Google Quantum AI and Collaborators}}},\ }\bibfield  {title} {\bibinfo {title} {Quantum error correction below the surface code threshold},\ }\bibfield  {journal} {\bibinfo  {journal} {Nature}\ }\href {https://doi.org/10.1038/s41586-024-08449-y} {10.1038/s41586-024-08449-y} (\bibinfo {year} {2024})\BibitemShut {NoStop}%
\bibitem [{\citenamefont {Bluvstein}\ \emph {et~al.}(2024)\citenamefont {Bluvstein} \emph {et~al.}}]{bluvsteinLogicalQuantumProcessor2024}%
  \BibitemOpen
  \bibfield  {author} {\bibinfo {author} {\bibfnamefont {D.}~\bibnamefont {Bluvstein}} \emph {et~al.},\ }\bibfield  {title} {\bibinfo {title} {Logical quantum processor based on reconfigurable atom arrays},\ }\href {https://doi.org/10.1038/s41586-023-06927-3} {\bibfield  {journal} {\bibinfo  {journal} {Nature}\ }\textbf {\bibinfo {volume} {626}},\ \bibinfo {pages} {58} (\bibinfo {year} {2024})}\BibitemShut {NoStop}%
\bibitem [{\citenamefont {Rodriguez}\ \emph {et~al.}(2024)\citenamefont {Rodriguez} \emph {et~al.}}]{rodriguezExperimentalDemonstrationLogical2024}%
  \BibitemOpen
  \bibfield  {author} {\bibinfo {author} {\bibfnamefont {P.~S.}\ \bibnamefont {Rodriguez}} \emph {et~al.},\ }\href {https://doi.org/10.48550/arXiv.2412.15165} {\bibinfo {title} {Experimental {{Demonstration}} of {{Logical Magic State Distillation}}}} (\bibinfo {year} {2024}),\ \Eprint {https://arxiv.org/abs/2412.15165} {arXiv:2412.15165 [quant-ph]} \BibitemShut {NoStop}%
\bibitem [{\citenamefont {Reichardt}\ \emph {et~al.}(2024)\citenamefont {Reichardt} \emph {et~al.}}]{reichardtDemonstrationQuantumComputation2024}%
  \BibitemOpen
  \bibfield  {author} {\bibinfo {author} {\bibfnamefont {B.~W.}\ \bibnamefont {Reichardt}} \emph {et~al.},\ }\href {https://doi.org/10.48550/arXiv.2409.04628} {\bibinfo {title} {Demonstration of quantum computation and error correction with a tesseract code}} (\bibinfo {year} {2024}),\ \Eprint {https://arxiv.org/abs/2409.04628} {arXiv:2409.04628 [quant-ph]} \BibitemShut {NoStop}%
\bibitem [{\citenamefont {Campbell}(2024)}]{campbell2024series}%
  \BibitemOpen
  \bibfield  {author} {\bibinfo {author} {\bibfnamefont {E.}~\bibnamefont {Campbell}},\ }\bibfield  {title} {\bibinfo {title} {A series of fast-paced advances in quantum error correction},\ }\href@noop {} {\bibfield  {journal} {\bibinfo  {journal} {Nature Reviews Physics}\ }\textbf {\bibinfo {volume} {6}},\ \bibinfo {pages} {160} (\bibinfo {year} {2024})}\BibitemShut {NoStop}%
\bibitem [{\citenamefont {Campbell}(2022)}]{campbellEarlyFaulttolerantSimulations2022}%
  \BibitemOpen
  \bibfield  {author} {\bibinfo {author} {\bibfnamefont {E.~T.}\ \bibnamefont {Campbell}},\ }\bibfield  {title} {\bibinfo {title} {Early fault-tolerant simulations of the {{Hubbard}} model},\ }\href {https://doi.org/10.1088/2058-9565/ac3110} {\bibfield  {journal} {\bibinfo  {journal} {Quantum Sci. Technol.}\ }\textbf {\bibinfo {volume} {7}},\ \bibinfo {pages} {015007} (\bibinfo {year} {2022})},\ \Eprint {https://arxiv.org/abs/2012.09238} {arXiv:2012.09238 [quant-ph]} \BibitemShut {NoStop}%
\bibitem [{\citenamefont {Katabarwa}\ \emph {et~al.}(2024)\citenamefont {Katabarwa} \emph {et~al.}}]{katabarwaEarlyFaultTolerantQuantum2024}%
  \BibitemOpen
  \bibfield  {author} {\bibinfo {author} {\bibfnamefont {A.}~\bibnamefont {Katabarwa}} \emph {et~al.},\ }\bibfield  {title} {\bibinfo {title} {Early {{Fault-Tolerant Quantum Computing}}},\ }\href {https://doi.org/10.1103/PRXQuantum.5.020101} {\bibfield  {journal} {\bibinfo  {journal} {PRX Quantum}\ }\textbf {\bibinfo {volume} {5}},\ \bibinfo {pages} {020101} (\bibinfo {year} {2024})}\BibitemShut {NoStop}%
\bibitem [{\citenamefont {Toshio}\ \emph {et~al.}(2024)\citenamefont {Toshio}, \citenamefont {Akahoshi}, \citenamefont {Fujisaki}, \citenamefont {Oshima}, \citenamefont {Sato},\ and\ \citenamefont {Fujii}}]{toshioPracticalQuantumAdvantage2024}%
  \BibitemOpen
  \bibfield  {author} {\bibinfo {author} {\bibfnamefont {R.}~\bibnamefont {Toshio}}, \bibinfo {author} {\bibfnamefont {Y.}~\bibnamefont {Akahoshi}}, \bibinfo {author} {\bibfnamefont {J.}~\bibnamefont {Fujisaki}}, \bibinfo {author} {\bibfnamefont {H.}~\bibnamefont {Oshima}}, \bibinfo {author} {\bibfnamefont {S.}~\bibnamefont {Sato}},\ and\ \bibinfo {author} {\bibfnamefont {K.}~\bibnamefont {Fujii}},\ }\href {https://arxiv.org/abs/2408.14848} {\bibinfo {title} {Practical quantum advantage on partially fault-tolerant quantum computer}} (\bibinfo {year} {2024}),\ \Eprint {https://arxiv.org/abs/2408.14848} {arXiv:2408.14848 [quant-ph]} \BibitemShut {NoStop}%
\bibitem [{\citenamefont {Preskill}(2025)}]{preskillBeyondNISQMegaquop2025}%
  \BibitemOpen
  \bibfield  {author} {\bibinfo {author} {\bibfnamefont {J.}~\bibnamefont {Preskill}},\ }\bibfield  {title} {\bibinfo {title} {Beyond nisq: The megaquop machine},\ }\bibfield  {journal} {\bibinfo  {journal} {ACM Transactions on Quantum Computing}\ }\href {https://doi.org/10.1145/3723153} {10.1145/3723153} (\bibinfo {year} {2025})\BibitemShut {NoStop}%
\bibitem [{\citenamefont {Gottesman}(1998)}]{gottesmanHeisenbergRepresentationQuantum1998}%
  \BibitemOpen
  \bibfield  {author} {\bibinfo {author} {\bibfnamefont {D.}~\bibnamefont {Gottesman}},\ }\href@noop {} {\bibinfo {title} {The {{Heisenberg Representation}} of {{Quantum Computers}}}} (\bibinfo {year} {1998}),\ \Eprint {https://arxiv.org/abs/quant-ph/9807006} {arXiv:quant-ph/9807006} \BibitemShut {NoStop}%
\bibitem [{\citenamefont {Bravyi}\ and\ \citenamefont {Kitaev}(2005)}]{bravyiUniversalQuantumComputation2005}%
  \BibitemOpen
  \bibfield  {author} {\bibinfo {author} {\bibfnamefont {S.}~\bibnamefont {Bravyi}}\ and\ \bibinfo {author} {\bibfnamefont {A.}~\bibnamefont {Kitaev}},\ }\bibfield  {title} {\bibinfo {title} {Universal quantum computation with ideal {{Clifford}} gates and noisy ancillas},\ }\href {https://doi.org/10.1103/PhysRevA.71.022316} {\bibfield  {journal} {\bibinfo  {journal} {Phys. Rev. A}\ }\textbf {\bibinfo {volume} {71}},\ \bibinfo {pages} {022316} (\bibinfo {year} {2005})}\BibitemShut {NoStop}%
\bibitem [{\citenamefont {Litinski}(2019{\natexlab{a}})}]{litinskiMagicStateDistillation2019}%
  \BibitemOpen
  \bibfield  {author} {\bibinfo {author} {\bibfnamefont {D.}~\bibnamefont {Litinski}},\ }\bibfield  {title} {\bibinfo {title} {Magic {{State Distillation}}: {{Not}} as {{Costly}} as {{You Think}}},\ }\href {https://doi.org/10.22331/q-2019-12-02-205} {\bibfield  {journal} {\bibinfo  {journal} {Quantum}\ }\textbf {\bibinfo {volume} {3}},\ \bibinfo {pages} {205} (\bibinfo {year} {2019}{\natexlab{a}})}\BibitemShut {NoStop}%
\bibitem [{\citenamefont {Gidney}\ and\ \citenamefont {Fowler}(2019)}]{gidney2019efficient}%
  \BibitemOpen
  \bibfield  {author} {\bibinfo {author} {\bibfnamefont {C.}~\bibnamefont {Gidney}}\ and\ \bibinfo {author} {\bibfnamefont {A.~G.}\ \bibnamefont {Fowler}},\ }\bibfield  {title} {\bibinfo {title} {Efficient magic state factories with a catalyzed $| \mathrm{CCZ} \rangle $ to $2| \mathrm{T} \rangle $ transformation},\ }\href@noop {} {\bibfield  {journal} {\bibinfo  {journal} {Quantum}\ }\textbf {\bibinfo {volume} {3}},\ \bibinfo {pages} {135} (\bibinfo {year} {2019})}\BibitemShut {NoStop}%
\bibitem [{\citenamefont {Wills}\ \emph {et~al.}(2024)\citenamefont {Wills}, \citenamefont {Hsieh},\ and\ \citenamefont {Yamasaki}}]{MSD24}%
  \BibitemOpen
  \bibfield  {author} {\bibinfo {author} {\bibfnamefont {A.}~\bibnamefont {Wills}}, \bibinfo {author} {\bibfnamefont {M.-H.}\ \bibnamefont {Hsieh}},\ and\ \bibinfo {author} {\bibfnamefont {H.}~\bibnamefont {Yamasaki}},\ }\href {https://doi.org/10.48550/arXiv.2408.07764} {\bibinfo {title} {Constant-{Overhead Magic State Distillation}}} (\bibinfo {year} {2024}),\ \Eprint {https://arxiv.org/abs/2408.07764} {arXiv:2408.07764 [quant-ph]} \BibitemShut {NoStop}%
\bibitem [{\citenamefont {O'Gorman}\ and\ \citenamefont {Campbell}(2017)}]{ogormanQuantumComputationRealistic2017}%
  \BibitemOpen
  \bibfield  {author} {\bibinfo {author} {\bibfnamefont {J.}~\bibnamefont {O'Gorman}}\ and\ \bibinfo {author} {\bibfnamefont {E.~T.}\ \bibnamefont {Campbell}},\ }\bibfield  {title} {\bibinfo {title} {Quantum computation with realistic magic-state factories},\ }\href {https://doi.org/10.1103/PhysRevA.95.032338} {\bibfield  {journal} {\bibinfo  {journal} {Phys. Rev. A}\ }\textbf {\bibinfo {volume} {95}},\ \bibinfo {pages} {032338} (\bibinfo {year} {2017})}\BibitemShut {NoStop}%
\bibitem [{\citenamefont {Lee}\ \emph {et~al.}(2021)\citenamefont {Lee}, \citenamefont {Berry}, \citenamefont {Gidney}, \citenamefont {Huggins}, \citenamefont {McClean}, \citenamefont {Wiebe},\ and\ \citenamefont {Babbush}}]{TensorHypercontraction}%
  \BibitemOpen
  \bibfield  {author} {\bibinfo {author} {\bibfnamefont {J.}~\bibnamefont {Lee}}, \bibinfo {author} {\bibfnamefont {D.~W.}\ \bibnamefont {Berry}}, \bibinfo {author} {\bibfnamefont {C.}~\bibnamefont {Gidney}}, \bibinfo {author} {\bibfnamefont {W.~J.}\ \bibnamefont {Huggins}}, \bibinfo {author} {\bibfnamefont {J.~R.}\ \bibnamefont {McClean}}, \bibinfo {author} {\bibfnamefont {N.}~\bibnamefont {Wiebe}},\ and\ \bibinfo {author} {\bibfnamefont {R.}~\bibnamefont {Babbush}},\ }\bibfield  {title} {\bibinfo {title} {Even more efficient quantum computations of chemistry through tensor hypercontraction},\ }\href {https://doi.org/10.1103/PRXQuantum.2.030305} {\bibfield  {journal} {\bibinfo  {journal} {PRX Quantum}\ }\textbf {\bibinfo {volume} {2}},\ \bibinfo {pages} {030305} (\bibinfo {year} {2021})}\BibitemShut {NoStop}%
\bibitem [{\citenamefont {Gidney}(2025)}]{gidney2025factor}%
  \BibitemOpen
  \bibfield  {author} {\bibinfo {author} {\bibfnamefont {C.}~\bibnamefont {Gidney}},\ }\href {https://arxiv.org/abs/2505.15917} {\bibinfo {title} {How to factor 2048 bit rsa integers with less than a million noisy qubits}} (\bibinfo {year} {2025}),\ \Eprint {https://arxiv.org/abs/2505.15917} {arXiv:2505.15917 [quant-ph]} \BibitemShut {NoStop}%
\bibitem [{\citenamefont {Georges}\ \emph {et~al.}(2025)\citenamefont {Georges}, \citenamefont {Bothe}, \citenamefont {S{\"u}nderhauf}, \citenamefont {Berntson}, \citenamefont {Izs{\'a}k},\ and\ \citenamefont {Ivanov}}]{georges2025quantum}%
  \BibitemOpen
  \bibfield  {author} {\bibinfo {author} {\bibfnamefont {T.~N.}\ \bibnamefont {Georges}}, \bibinfo {author} {\bibfnamefont {M.}~\bibnamefont {Bothe}}, \bibinfo {author} {\bibfnamefont {C.}~\bibnamefont {S{\"u}nderhauf}}, \bibinfo {author} {\bibfnamefont {B.~K.}\ \bibnamefont {Berntson}}, \bibinfo {author} {\bibfnamefont {R.}~\bibnamefont {Izs{\'a}k}},\ and\ \bibinfo {author} {\bibfnamefont {A.~V.}\ \bibnamefont {Ivanov}},\ }\bibfield  {title} {\bibinfo {title} {Quantum simulations of chemistry in first quantization with any basis set},\ }\href@noop {} {\bibfield  {journal} {\bibinfo  {journal} {npj Quantum Information}\ }\textbf {\bibinfo {volume} {11}},\ \bibinfo {pages} {55} (\bibinfo {year} {2025})}\BibitemShut {NoStop}%
\bibitem [{\citenamefont {Peruzzo}\ \emph {et~al.}(2014)\citenamefont {Peruzzo} \emph {et~al.}}]{peruzzoVariationalEigenvalueSolver2014}%
  \BibitemOpen
  \bibfield  {author} {\bibinfo {author} {\bibfnamefont {A.}~\bibnamefont {Peruzzo}} \emph {et~al.},\ }\bibfield  {title} {\bibinfo {title} {A variational eigenvalue solver on a photonic quantum processor},\ }\href {https://doi.org/10.1038/ncomms5213} {\bibfield  {journal} {\bibinfo  {journal} {Nat Commun}\ }\textbf {\bibinfo {volume} {5}},\ \bibinfo {pages} {4213} (\bibinfo {year} {2014})}\BibitemShut {NoStop}%
\bibitem [{\citenamefont {Somma}(2019)}]{sommaQuantumEigenvalueEstimation2019}%
  \BibitemOpen
  \bibfield  {author} {\bibinfo {author} {\bibfnamefont {R.~D.}\ \bibnamefont {Somma}},\ }\bibfield  {title} {\bibinfo {title} {Quantum eigenvalue estimation via time series analysis},\ }\href {https://doi.org/10.1088/1367-2630/ab5c60} {\bibfield  {journal} {\bibinfo  {journal} {New J. Phys.}\ }\textbf {\bibinfo {volume} {21}},\ \bibinfo {pages} {123025} (\bibinfo {year} {2019})}\BibitemShut {NoStop}%
\bibitem [{\citenamefont {Lin}\ and\ \citenamefont {Tong}(2022)}]{linHeisenbergLimitedGroundStateEnergy2022}%
  \BibitemOpen
  \bibfield  {author} {\bibinfo {author} {\bibfnamefont {L.}~\bibnamefont {Lin}}\ and\ \bibinfo {author} {\bibfnamefont {Y.}~\bibnamefont {Tong}},\ }\bibfield  {title} {\bibinfo {title} {Heisenberg-{{Limited Ground-State Energy Estimation}} for {{Early Fault-Tolerant Quantum Computers}}},\ }\href {https://doi.org/10.1103/PRXQuantum.3.010318} {\bibfield  {journal} {\bibinfo  {journal} {PRX Quantum}\ }\textbf {\bibinfo {volume} {3}},\ \bibinfo {pages} {010318} (\bibinfo {year} {2022})}\BibitemShut {NoStop}%
\bibitem [{\citenamefont {Wan}\ \emph {et~al.}(2022)\citenamefont {Wan}, \citenamefont {Berta},\ and\ \citenamefont {Campbell}}]{wanRandomizedQuantumAlgorithm2022}%
  \BibitemOpen
  \bibfield  {author} {\bibinfo {author} {\bibfnamefont {K.}~\bibnamefont {Wan}}, \bibinfo {author} {\bibfnamefont {M.}~\bibnamefont {Berta}},\ and\ \bibinfo {author} {\bibfnamefont {E.~T.}\ \bibnamefont {Campbell}},\ }\bibfield  {title} {\bibinfo {title} {Randomized {{Quantum Algorithm}} for {{Statistical Phase Estimation}}},\ }\href {https://doi.org/10.1103/PhysRevLett.129.030503} {\bibfield  {journal} {\bibinfo  {journal} {Phys. Rev. Lett.}\ }\textbf {\bibinfo {volume} {129}},\ \bibinfo {pages} {030503} (\bibinfo {year} {2022})}\BibitemShut {NoStop}%
\bibitem [{\citenamefont {Blunt}\ \emph {et~al.}(2023)\citenamefont {Blunt} \emph {et~al.}}]{bluntStatisticalPhaseEstimation2023}%
  \BibitemOpen
  \bibfield  {author} {\bibinfo {author} {\bibfnamefont {N.~S.}\ \bibnamefont {Blunt}} \emph {et~al.},\ }\bibfield  {title} {\bibinfo {title} {Statistical phase estimation and error mitigation on a superconducting quantum processor},\ }\href {https://doi.org/10.1103/PRXQuantum.4.040341} {\bibfield  {journal} {\bibinfo  {journal} {PRX Quantum}\ }\textbf {\bibinfo {volume} {4}},\ \bibinfo {pages} {040341} (\bibinfo {year} {2023})}\BibitemShut {NoStop}%
\bibitem [{\citenamefont {G{\"u}nther}\ \emph {et~al.}(2025)\citenamefont {G{\"u}nther}, \citenamefont {Witteveen}, \citenamefont {Schmidhuber}, \citenamefont {Miller}, \citenamefont {Christandl},\ and\ \citenamefont {Harrow}}]{gunther2025phase}%
  \BibitemOpen
  \bibfield  {author} {\bibinfo {author} {\bibfnamefont {J.}~\bibnamefont {G{\"u}nther}}, \bibinfo {author} {\bibfnamefont {F.}~\bibnamefont {Witteveen}}, \bibinfo {author} {\bibfnamefont {A.}~\bibnamefont {Schmidhuber}}, \bibinfo {author} {\bibfnamefont {M.}~\bibnamefont {Miller}}, \bibinfo {author} {\bibfnamefont {M.}~\bibnamefont {Christandl}},\ and\ \bibinfo {author} {\bibfnamefont {A.}~\bibnamefont {Harrow}},\ }\bibfield  {title} {\bibinfo {title} {Phase estimation with partially randomized time evolution},\ }\href@noop {} {\bibfield  {journal} {\bibinfo  {journal} {arXiv preprint arXiv:2503.05647}\ } (\bibinfo {year} {2025})}\BibitemShut {NoStop}%
\bibitem [{\citenamefont {Temme}\ \emph {et~al.}(2017)\citenamefont {Temme}, \citenamefont {Bravyi},\ and\ \citenamefont {Gambetta}}]{temmeErrorMitigationShortDepth2017}%
  \BibitemOpen
  \bibfield  {author} {\bibinfo {author} {\bibfnamefont {K.}~\bibnamefont {Temme}}, \bibinfo {author} {\bibfnamefont {S.}~\bibnamefont {Bravyi}},\ and\ \bibinfo {author} {\bibfnamefont {J.~M.}\ \bibnamefont {Gambetta}},\ }\bibfield  {title} {\bibinfo {title} {Error {{Mitigation}} for {{Short-Depth Quantum Circuits}}},\ }\href {https://doi.org/10.1103/PhysRevLett.119.180509} {\bibfield  {journal} {\bibinfo  {journal} {Phys. Rev. Lett.}\ }\textbf {\bibinfo {volume} {119}},\ \bibinfo {pages} {180509} (\bibinfo {year} {2017})}\BibitemShut {NoStop}%
\bibitem [{\citenamefont {Cai}\ \emph {et~al.}(2023)\citenamefont {Cai} \emph {et~al.}}]{caiQuantumErrorMitigation2023}%
  \BibitemOpen
  \bibfield  {author} {\bibinfo {author} {\bibfnamefont {Z.}~\bibnamefont {Cai}} \emph {et~al.},\ }\bibfield  {title} {\bibinfo {title} {Quantum {{Error Mitigation}}},\ }\href {https://doi.org/10.1103/RevModPhys.95.045005} {\bibfield  {journal} {\bibinfo  {journal} {Rev. Mod. Phys.}\ }\textbf {\bibinfo {volume} {95}},\ \bibinfo {pages} {045005} (\bibinfo {year} {2023})}\BibitemShut {NoStop}%
\bibitem [{\citenamefont {Piveteau}\ \emph {et~al.}(2021)\citenamefont {Piveteau} \emph {et~al.}}]{piveteauErrorMitigationUniversal2021}%
  \BibitemOpen
  \bibfield  {author} {\bibinfo {author} {\bibfnamefont {C.}~\bibnamefont {Piveteau}} \emph {et~al.},\ }\bibfield  {title} {\bibinfo {title} {Error {{Mitigation}} for {{Universal Gates}} on {{Encoded Qubits}}},\ }\href {https://doi.org/10.1103/PhysRevLett.127.200505} {\bibfield  {journal} {\bibinfo  {journal} {Phys. Rev. Lett.}\ }\textbf {\bibinfo {volume} {127}},\ \bibinfo {pages} {200505} (\bibinfo {year} {2021})}\BibitemShut {NoStop}%
\bibitem [{\citenamefont {Suzuki}\ \emph {et~al.}(2022)\citenamefont {Suzuki} \emph {et~al.}}]{suzukiQuantumErrorMitigation2022}%
  \BibitemOpen
  \bibfield  {author} {\bibinfo {author} {\bibfnamefont {Y.}~\bibnamefont {Suzuki}} \emph {et~al.},\ }\bibfield  {title} {\bibinfo {title} {Quantum {{Error Mitigation}} as a {{Universal Error Reduction Technique}}: {{Applications}} from the {{NISQ}} to the {{Fault-Tolerant Quantum Computing Eras}}},\ }\href {https://doi.org/10.1103/PRXQuantum.3.010345} {\bibfield  {journal} {\bibinfo  {journal} {PRX Quantum}\ }\textbf {\bibinfo {volume} {3}},\ \bibinfo {pages} {010345} (\bibinfo {year} {2022})}\BibitemShut {NoStop}%
\bibitem [{\citenamefont {Ross}\ and\ \citenamefont {Selinger}(2016)}]{rossOptimalAncillafreeClifford+T2016}%
  \BibitemOpen
  \bibfield  {author} {\bibinfo {author} {\bibfnamefont {N.~J.}\ \bibnamefont {Ross}}\ and\ \bibinfo {author} {\bibfnamefont {P.}~\bibnamefont {Selinger}},\ }\href {https://doi.org/10.48550/arXiv.1403.2975} {\bibinfo {title} {Optimal ancilla-free {{Clifford}}+{{T}} approximation of z-rotations}} (\bibinfo {year} {2016}),\ \Eprint {https://arxiv.org/abs/1403.2975} {arXiv:1403.2975} \BibitemShut {NoStop}%
\bibitem [{\citenamefont {Veitch}\ \emph {et~al.}(2014)\citenamefont {Veitch} \emph {et~al.}}]{veitchResourceTheoryStabilizer2014}%
  \BibitemOpen
  \bibfield  {author} {\bibinfo {author} {\bibfnamefont {V.}~\bibnamefont {Veitch}} \emph {et~al.},\ }\bibfield  {title} {\bibinfo {title} {The resource theory of stabilizer quantum computation},\ }\href {https://doi.org/10.1088/1367-2630/16/1/013009} {\bibfield  {journal} {\bibinfo  {journal} {New J. Phys.}\ }\textbf {\bibinfo {volume} {16}},\ \bibinfo {pages} {013009} (\bibinfo {year} {2014})}\BibitemShut {NoStop}%
\bibitem [{\citenamefont {Howard}\ and\ \citenamefont {Campbell}(2017)}]{howardApplicationResourceTheory2017}%
  \BibitemOpen
  \bibfield  {author} {\bibinfo {author} {\bibfnamefont {M.}~\bibnamefont {Howard}}\ and\ \bibinfo {author} {\bibfnamefont {E.}~\bibnamefont {Campbell}},\ }\bibfield  {title} {\bibinfo {title} {Application of a {{Resource Theory}} for {{Magic States}} to {{Fault-Tolerant Quantum Computing}}},\ }\href {https://doi.org/10.1103/PhysRevLett.118.090501} {\bibfield  {journal} {\bibinfo  {journal} {Phys. Rev. Lett.}\ }\textbf {\bibinfo {volume} {118}},\ \bibinfo {pages} {090501} (\bibinfo {year} {2017})}\BibitemShut {NoStop}%
\bibitem [{\citenamefont {Seddon}\ \emph {et~al.}(2021)\citenamefont {Seddon} \emph {et~al.}}]{seddonQuantifyingQuantumSpeedups2021}%
  \BibitemOpen
  \bibfield  {author} {\bibinfo {author} {\bibfnamefont {J.~R.}\ \bibnamefont {Seddon}} \emph {et~al.},\ }\bibfield  {title} {\bibinfo {title} {Quantifying {{Quantum Speedups}}: {{Improved Classical Simulation From Tighter Magic Monotones}}},\ }\href {https://doi.org/10.1103/PRXQuantum.2.010345} {\bibfield  {journal} {\bibinfo  {journal} {PRX Quantum}\ }\textbf {\bibinfo {volume} {2}},\ \bibinfo {pages} {010345} (\bibinfo {year} {2021})}\BibitemShut {NoStop}%
\bibitem [{\citenamefont {Seddon}\ and\ \citenamefont {Campbell}(2019)}]{seddonQuantifyingMagicMultiqubit2019}%
  \BibitemOpen
  \bibfield  {author} {\bibinfo {author} {\bibfnamefont {J.~R.}\ \bibnamefont {Seddon}}\ and\ \bibinfo {author} {\bibfnamefont {E.~T.}\ \bibnamefont {Campbell}},\ }\bibfield  {title} {\bibinfo {title} {Quantifying magic for multi-qubit operations},\ }\href {https://doi.org/10.1098/rspa.2019.0251} {\bibfield  {journal} {\bibinfo  {journal} {Proc. R. Soc. A.}\ }\textbf {\bibinfo {volume} {475}},\ \bibinfo {pages} {20190251} (\bibinfo {year} {2019})}\BibitemShut {NoStop}%
\bibitem [{\citenamefont {Bravyi}\ \emph {et~al.}(2019)\citenamefont {Bravyi} \emph {et~al.}}]{bravyiSimulationQuantumCircuits2019}%
  \BibitemOpen
  \bibfield  {author} {\bibinfo {author} {\bibfnamefont {S.}~\bibnamefont {Bravyi}} \emph {et~al.},\ }\bibfield  {title} {\bibinfo {title} {Simulation of quantum circuits by low-rank stabilizer decompositions},\ }\href {https://doi.org/10.22331/q-2019-09-02-181} {\bibfield  {journal} {\bibinfo  {journal} {Quantum}\ }\textbf {\bibinfo {volume} {3}},\ \bibinfo {pages} {181} (\bibinfo {year} {2019})}\BibitemShut {NoStop}%
\bibitem [{\citenamefont {Koczor}\ \emph {et~al.}(2024)\citenamefont {Koczor}, \citenamefont {Morton},\ and\ \citenamefont {Benjamin}}]{koczorProbabilisticInterpolationQuantum2024}%
  \BibitemOpen
  \bibfield  {author} {\bibinfo {author} {\bibfnamefont {B.}~\bibnamefont {Koczor}}, \bibinfo {author} {\bibfnamefont {J.~J.~L.}\ \bibnamefont {Morton}},\ and\ \bibinfo {author} {\bibfnamefont {S.~C.}\ \bibnamefont {Benjamin}},\ }\bibfield  {title} {\bibinfo {title} {Probabilistic {{Interpolation}} of {{Quantum Rotation Angles}}},\ }\href {https://doi.org/10.1103/PhysRevLett.132.130602} {\bibfield  {journal} {\bibinfo  {journal} {Phys. Rev. Lett.}\ }\textbf {\bibinfo {volume} {132}},\ \bibinfo {pages} {130602} (\bibinfo {year} {2024})}\BibitemShut {NoStop}%
\bibitem [{\citenamefont {Pashayan}\ \emph {et~al.}(2015)\citenamefont {Pashayan}, \citenamefont {Wallman},\ and\ \citenamefont {Bartlett}}]{pashayanEstimatingOutcomeProbabilities2015}%
  \BibitemOpen
  \bibfield  {author} {\bibinfo {author} {\bibfnamefont {H.}~\bibnamefont {Pashayan}}, \bibinfo {author} {\bibfnamefont {J.~J.}\ \bibnamefont {Wallman}},\ and\ \bibinfo {author} {\bibfnamefont {S.~D.}\ \bibnamefont {Bartlett}},\ }\bibfield  {title} {\bibinfo {title} {Estimating {{Outcome Probabilities}} of {{Quantum Circuits Using Quasiprobabilities}}},\ }\href {https://doi.org/10.1103/PhysRevLett.115.070501} {\bibfield  {journal} {\bibinfo  {journal} {Physical Review Letters}\ }\textbf {\bibinfo {volume} {115}},\ \bibinfo {pages} {070501} (\bibinfo {year} {2015})}\BibitemShut {NoStop}%
\bibitem [{\citenamefont {Gottesman}\ and\ \citenamefont {Chuang}(1999)}]{gottesmanDemonstratingViabilityUniversal1999}%
  \BibitemOpen
  \bibfield  {author} {\bibinfo {author} {\bibfnamefont {D.}~\bibnamefont {Gottesman}}\ and\ \bibinfo {author} {\bibfnamefont {I.~L.}\ \bibnamefont {Chuang}},\ }\bibfield  {title} {\bibinfo {title} {Demonstrating the viability of universal quantum computation using teleportation and single-qubit operations},\ }\href {https://doi.org/10.1038/46503} {\bibfield  {journal} {\bibinfo  {journal} {Nature}\ }\textbf {\bibinfo {volume} {402}},\ \bibinfo {pages} {390} (\bibinfo {year} {1999})}\BibitemShut {NoStop}%
\bibitem [{\citenamefont {Campbell}\ and\ \citenamefont {O'Gorman}(2016)}]{campbellEfficientMagicState2016}%
  \BibitemOpen
  \bibfield  {author} {\bibinfo {author} {\bibfnamefont {E.~T.}\ \bibnamefont {Campbell}}\ and\ \bibinfo {author} {\bibfnamefont {J.}~\bibnamefont {O'Gorman}},\ }\bibfield  {title} {\bibinfo {title} {An efficient magic state approach to small angle rotations},\ }\href {https://doi.org/10.1088/2058-9565/1/1/015007} {\bibfield  {journal} {\bibinfo  {journal} {Quantum Sci. Technol.}\ }\textbf {\bibinfo {volume} {1}},\ \bibinfo {pages} {015007} (\bibinfo {year} {2016})},\ \Eprint {https://arxiv.org/abs/1603.04230} {arXiv:1603.04230 [quant-ph]} \BibitemShut {NoStop}%
\bibitem [{\citenamefont {Li}(2015)}]{liMagicStatesFidelity2015}%
  \BibitemOpen
  \bibfield  {author} {\bibinfo {author} {\bibfnamefont {Y.}~\bibnamefont {Li}},\ }\bibfield  {title} {\bibinfo {title} {A magic state's fidelity can be superior to the operations that created it},\ }\href {https://doi.org/10.1088/1367-2630/17/2/023037} {\bibfield  {journal} {\bibinfo  {journal} {New J. Phys.}\ }\textbf {\bibinfo {volume} {17}},\ \bibinfo {pages} {023037} (\bibinfo {year} {2015})}\BibitemShut {NoStop}%
\bibitem [{\citenamefont {Hubbard}\ and\ \citenamefont {Flowers}(1963)}]{hubbardElectronCorr1963}%
  \BibitemOpen
  \bibfield  {author} {\bibinfo {author} {\bibfnamefont {J.}~\bibnamefont {Hubbard}}\ and\ \bibinfo {author} {\bibfnamefont {B.~H.}\ \bibnamefont {Flowers}},\ }\bibfield  {title} {\bibinfo {title} {Electron correlations in narrow energy bands},\ }\href {https://doi.org/10.1098/rspa.1963.0204} {\bibfield  {journal} {\bibinfo  {journal} {Proceedings of the Royal Society of London. Series A. Mathematical and Physical Sciences}\ }\textbf {\bibinfo {volume} {276}},\ \bibinfo {pages} {238} (\bibinfo {year} {1963})}\BibitemShut {NoStop}%
\bibitem [{Hub(2013)}]{HubbardModelHalf2013}%
  \BibitemOpen
  \bibfield  {title} {\bibinfo {title} {The {{Hubbard}} model at half a century},\ }\href {https://doi.org/10.1038/nphys2759} {\bibfield  {journal} {\bibinfo  {journal} {Nature Phys}\ }\textbf {\bibinfo {volume} {9}},\ \bibinfo {pages} {523} (\bibinfo {year} {2013})}\BibitemShut {NoStop}%
\bibitem [{\citenamefont {Cade}\ \emph {et~al.}(2020)\citenamefont {Cade} \emph {et~al.}}]{cadeStrategiesSolvingFermiHubbard2020}%
  \BibitemOpen
  \bibfield  {author} {\bibinfo {author} {\bibfnamefont {C.}~\bibnamefont {Cade}} \emph {et~al.},\ }\bibfield  {title} {\bibinfo {title} {Strategies for solving the {{Fermi-Hubbard}} model on near-term quantum computers},\ }\href {https://doi.org/10.1103/PhysRevB.102.235122} {\bibfield  {journal} {\bibinfo  {journal} {Phys. Rev. B}\ }\textbf {\bibinfo {volume} {102}},\ \bibinfo {pages} {235122} (\bibinfo {year} {2020})}\BibitemShut {NoStop}%
\bibitem [{\citenamefont {Dagotto}(1994)}]{dagottoCorrelatedElectronsHightemperature1994}%
  \BibitemOpen
  \bibfield  {author} {\bibinfo {author} {\bibfnamefont {E.}~\bibnamefont {Dagotto}},\ }\bibfield  {title} {\bibinfo {title} {Correlated electrons in high-temperature superconductors},\ }\href {https://doi.org/10.1103/RevModPhys.66.763} {\bibfield  {journal} {\bibinfo  {journal} {Rev. Mod. Phys.}\ }\textbf {\bibinfo {volume} {66}},\ \bibinfo {pages} {763} (\bibinfo {year} {1994})}\BibitemShut {NoStop}%
\bibitem [{\citenamefont {Imada}\ \emph {et~al.}(1998)\citenamefont {Imada}, \citenamefont {Fujimori},\ and\ \citenamefont {Tokura}}]{imadaMetalinsulatortransitions}%
  \BibitemOpen
  \bibfield  {author} {\bibinfo {author} {\bibfnamefont {M.}~\bibnamefont {Imada}}, \bibinfo {author} {\bibfnamefont {A.}~\bibnamefont {Fujimori}},\ and\ \bibinfo {author} {\bibfnamefont {Y.}~\bibnamefont {Tokura}},\ }\bibfield  {title} {\bibinfo {title} {Metal-insulator transitions},\ }\href {https://doi.org/10.1103/RevModPhys.70.1039} {\bibfield  {journal} {\bibinfo  {journal} {Rev. Mod. Phys.}\ }\textbf {\bibinfo {volume} {70}},\ \bibinfo {pages} {1039} (\bibinfo {year} {1998})}\BibitemShut {NoStop}%
\bibitem [{\citenamefont {Babbush}\ \emph {et~al.}(2021)\citenamefont {Babbush}, \citenamefont {McClean}, \citenamefont {Newman}, \citenamefont {Gidney}, \citenamefont {Boixo},\ and\ \citenamefont {Neven}}]{babbushBeyondQuadraticSpeedups2021}%
  \BibitemOpen
  \bibfield  {author} {\bibinfo {author} {\bibfnamefont {R.}~\bibnamefont {Babbush}}, \bibinfo {author} {\bibfnamefont {J.~R.}\ \bibnamefont {McClean}}, \bibinfo {author} {\bibfnamefont {M.}~\bibnamefont {Newman}}, \bibinfo {author} {\bibfnamefont {C.}~\bibnamefont {Gidney}}, \bibinfo {author} {\bibfnamefont {S.}~\bibnamefont {Boixo}},\ and\ \bibinfo {author} {\bibfnamefont {H.}~\bibnamefont {Neven}},\ }\bibfield  {title} {\bibinfo {title} {Focus beyond quadratic speedups for error-corrected quantum advantage},\ }\href {https://doi.org/10.1103/PRXQuantum.2.010103} {\bibfield  {journal} {\bibinfo  {journal} {PRX Quantum}\ }\textbf {\bibinfo {volume} {2}},\ \bibinfo {pages} {010103} (\bibinfo {year} {2021})}\BibitemShut {NoStop}%
\bibitem [{\citenamefont {Litinski}(2019{\natexlab{b}})}]{litinskiGameSurfaceCodes2019}%
  \BibitemOpen
  \bibfield  {author} {\bibinfo {author} {\bibfnamefont {D.}~\bibnamefont {Litinski}},\ }\bibfield  {title} {\bibinfo {title} {A {{Game}} of {{Surface Codes}}: {{Large-Scale Quantum Computing}} with {{Lattice Surgery}}},\ }\href {https://doi.org/10.22331/q-2019-03-05-128} {\bibfield  {journal} {\bibinfo  {journal} {Quantum}\ }\textbf {\bibinfo {volume} {3}},\ \bibinfo {pages} {128} (\bibinfo {year} {2019}{\natexlab{b}})},\ \Eprint {https://arxiv.org/abs/1808.02892} {arXiv:1808.02892 [cond-mat, physics:quant-ph]} \BibitemShut {NoStop}%
\bibitem [{\citenamefont {Endo}\ \emph {et~al.}(2018)\citenamefont {Endo}, \citenamefont {Benjamin},\ and\ \citenamefont {Li}}]{endoPracticalQuantumError2018}%
  \BibitemOpen
  \bibfield  {author} {\bibinfo {author} {\bibfnamefont {S.}~\bibnamefont {Endo}}, \bibinfo {author} {\bibfnamefont {S.~C.}\ \bibnamefont {Benjamin}},\ and\ \bibinfo {author} {\bibfnamefont {Y.}~\bibnamefont {Li}},\ }\bibfield  {title} {\bibinfo {title} {Practical {{Quantum Error Mitigation}} for {{Near-Future Applications}}},\ }\href {https://doi.org/10.1103/PhysRevX.8.031027} {\bibfield  {journal} {\bibinfo  {journal} {Phys. Rev. X}\ }\textbf {\bibinfo {volume} {8}},\ \bibinfo {pages} {031027} (\bibinfo {year} {2018})}\BibitemShut {NoStop}%
\bibitem [{\citenamefont {Diamond}\ and\ \citenamefont {Boyd}(2016)}]{diamond2016cvxpy}%
  \BibitemOpen
  \bibfield  {author} {\bibinfo {author} {\bibfnamefont {S.}~\bibnamefont {Diamond}}\ and\ \bibinfo {author} {\bibfnamefont {S.}~\bibnamefont {Boyd}},\ }\bibfield  {title} {\bibinfo {title} {{CVXPY}: {A} {P}ython-embedded modeling language for convex optimization},\ }\href@noop {} {\bibfield  {journal} {\bibinfo  {journal} {Journal of Machine Learning Research}\ }\textbf {\bibinfo {volume} {17}},\ \bibinfo {pages} {1} (\bibinfo {year} {2016})}\BibitemShut {NoStop}%
\bibitem [{\citenamefont {Agrawal}\ \emph {et~al.}(2018)\citenamefont {Agrawal}, \citenamefont {Verschueren}, \citenamefont {Diamond},\ and\ \citenamefont {Boyd}}]{agrawal2018rewriting}%
  \BibitemOpen
  \bibfield  {author} {\bibinfo {author} {\bibfnamefont {A.}~\bibnamefont {Agrawal}}, \bibinfo {author} {\bibfnamefont {R.}~\bibnamefont {Verschueren}}, \bibinfo {author} {\bibfnamefont {S.}~\bibnamefont {Diamond}},\ and\ \bibinfo {author} {\bibfnamefont {S.}~\bibnamefont {Boyd}},\ }\bibfield  {title} {\bibinfo {title} {A rewriting system for convex optimization problems},\ }\href@noop {} {\bibfield  {journal} {\bibinfo  {journal} {Journal of Control and Decision}\ }\textbf {\bibinfo {volume} {5}},\ \bibinfo {pages} {42} (\bibinfo {year} {2018})}\BibitemShut {NoStop}%
\bibitem [{Note1()}]{Note1}%
  \BibitemOpen
  \bibinfo {note} {Note that if $Z$ errors occur on both the $|T\rangle $ and $|T^\protect \frac {1}{2}\rangle $ states, then the overall effect on the $T^\protect \frac {1}{2}$ gate is $Z^2=I$, meaning this case is equivalent to no error occurring.}\BibitemShut {Stop}%
\bibitem [{\citenamefont {Kivlichan}\ \emph {et~al.}(2020)\citenamefont {Kivlichan} \emph {et~al.}}]{kivlichanImprovedFaultTolerantQuantum2020}%
  \BibitemOpen
  \bibfield  {author} {\bibinfo {author} {\bibfnamefont {I.~D.}\ \bibnamefont {Kivlichan}} \emph {et~al.},\ }\bibfield  {title} {\bibinfo {title} {Improved {{Fault-Tolerant Quantum Simulation}} of {{Condensed-Phase Correlated Electrons}} via {{Trotterization}}},\ }\href {https://doi.org/10.22331/q-2020-07-16-296} {\bibfield  {journal} {\bibinfo  {journal} {Quantum}\ }\textbf {\bibinfo {volume} {4}},\ \bibinfo {pages} {296} (\bibinfo {year} {2020})},\ \Eprint {https://arxiv.org/abs/1902.10673} {arXiv:1902.10673 [quant-ph]} \BibitemShut {NoStop}%
\bibitem [{\citenamefont {Campbell}(2017)}]{campbellShorterGateSequences2017}%
  \BibitemOpen
  \bibfield  {author} {\bibinfo {author} {\bibfnamefont {E.}~\bibnamefont {Campbell}},\ }\bibfield  {title} {\bibinfo {title} {Shorter gate sequences for quantum computing by mixing unitaries},\ }\href {https://doi.org/10.1103/PhysRevA.95.042306} {\bibfield  {journal} {\bibinfo  {journal} {Phys. Rev. A}\ }\textbf {\bibinfo {volume} {95}},\ \bibinfo {pages} {042306} (\bibinfo {year} {2017})}\BibitemShut {NoStop}%
\bibitem [{\citenamefont {Kliuchnikov}\ \emph {et~al.}(2023)\citenamefont {Kliuchnikov}, \citenamefont {Lauter}, \citenamefont {Minko}, \citenamefont {Paetznick},\ and\ \citenamefont {Petit}}]{kliuchnikovShorterQuantum2023}%
  \BibitemOpen
  \bibfield  {author} {\bibinfo {author} {\bibfnamefont {V.}~\bibnamefont {Kliuchnikov}}, \bibinfo {author} {\bibfnamefont {K.}~\bibnamefont {Lauter}}, \bibinfo {author} {\bibfnamefont {R.}~\bibnamefont {Minko}}, \bibinfo {author} {\bibfnamefont {A.}~\bibnamefont {Paetznick}},\ and\ \bibinfo {author} {\bibfnamefont {C.}~\bibnamefont {Petit}},\ }\bibfield  {title} {\bibinfo {title} {Shorter quantum circuits via single-qubit gate approximation},\ }\href {https://doi.org/10.22331/q-2023-12-18-1208} {\bibfield  {journal} {\bibinfo  {journal} {{Quantum}}\ }\textbf {\bibinfo {volume} {7}},\ \bibinfo {pages} {1208} (\bibinfo {year} {2023})}\BibitemShut {NoStop}%
\bibitem [{\citenamefont {Yoshioka}\ \emph {et~al.}(2025)\citenamefont {Yoshioka}, \citenamefont {Akibue}, \citenamefont {Morisaki}, \citenamefont {Morisaki}, \citenamefont {Tsubouchi},\ and\ \citenamefont {Suzuki}}]{yoshiokaErrorCrafting2025}%
  \BibitemOpen
  \bibfield  {author} {\bibinfo {author} {\bibfnamefont {N.}~\bibnamefont {Yoshioka}}, \bibinfo {author} {\bibfnamefont {S.}~\bibnamefont {Akibue}}, \bibinfo {author} {\bibfnamefont {H.}~\bibnamefont {Morisaki}}, \bibinfo {author} {\bibfnamefont {H.}~\bibnamefont {Morisaki}}, \bibinfo {author} {\bibfnamefont {K.}~\bibnamefont {Tsubouchi}},\ and\ \bibinfo {author} {\bibfnamefont {Y.}~\bibnamefont {Suzuki}},\ }\bibfield  {title} {\bibinfo {title} {Error crafting in mixed quantum gate synthesis},\ }\href {https://doi.org/10.1038/s41534-025-01032-x} {\bibfield  {journal} {\bibinfo  {journal} {npj Quantum Information}\ }\textbf {\bibinfo {volume} {11}},\ \bibinfo {pages} {95} (\bibinfo {year} {2025})}\BibitemShut {NoStop}%
\bibitem [{\citenamefont {Choi}\ \emph {et~al.}(2023)\citenamefont {Choi} \emph {et~al.}}]{choiFaultTolerantNonClifford2023}%
  \BibitemOpen
  \bibfield  {author} {\bibinfo {author} {\bibfnamefont {H.}~\bibnamefont {Choi}} \emph {et~al.},\ }\href {https://doi.org/10.48550/arXiv.2303.17380} {\bibinfo {title} {Fault {{Tolerant Non-Clifford State Preparation}} for {{Arbitrary Rotations}}}} (\bibinfo {year} {2023}),\ \Eprint {https://arxiv.org/abs/2303.17380} {arXiv:2303.17380 [quant-ph]} \BibitemShut {NoStop}%
\bibitem [{\citenamefont {Gidney}\ \emph {et~al.}(2024)\citenamefont {Gidney}, \citenamefont {Shutty},\ and\ \citenamefont {Jones}}]{gidneyMagicStateCultivation2024}%
  \BibitemOpen
  \bibfield  {author} {\bibinfo {author} {\bibfnamefont {C.}~\bibnamefont {Gidney}}, \bibinfo {author} {\bibfnamefont {N.}~\bibnamefont {Shutty}},\ and\ \bibinfo {author} {\bibfnamefont {C.}~\bibnamefont {Jones}},\ }\href {https://doi.org/10.48550/arXiv.2409.17595} {\bibinfo {title} {Magic state cultivation: Growing {{T}} states as cheap as {{CNOT}} gates}} (\bibinfo {year} {2024}),\ \Eprint {https://arxiv.org/abs/2409.17595} {arXiv:2409.17595 [quant-ph]} \BibitemShut {NoStop}%
\bibitem [{\citenamefont {Horsman}\ \emph {et~al.}(2012)\citenamefont {Horsman}, \citenamefont {Fowler}, \citenamefont {Devitt},\ and\ \citenamefont {Meter}}]{horsman2012surface}%
  \BibitemOpen
  \bibfield  {author} {\bibinfo {author} {\bibfnamefont {D.}~\bibnamefont {Horsman}}, \bibinfo {author} {\bibfnamefont {A.~G.}\ \bibnamefont {Fowler}}, \bibinfo {author} {\bibfnamefont {S.}~\bibnamefont {Devitt}},\ and\ \bibinfo {author} {\bibfnamefont {R.~V.}\ \bibnamefont {Meter}},\ }\bibfield  {title} {\bibinfo {title} {Surface code quantum computing by lattice surgery},\ }\href {https://doi.org/10.1088/1367-2630/14/12/123011} {\bibfield  {journal} {\bibinfo  {journal} {New Journal of Physics}\ }\textbf {\bibinfo {volume} {14}},\ \bibinfo {pages} {123011} (\bibinfo {year} {2012})}\BibitemShut {NoStop}%
\bibitem [{\citenamefont {Geh{\'{e}}r}\ \emph {et~al.}(2024)\citenamefont {Geh{\'{e}}r} \emph {et~al.}}]{geher2024error}%
  \BibitemOpen
  \bibfield  {author} {\bibinfo {author} {\bibfnamefont {G.~P.}\ \bibnamefont {Geh{\'{e}}r}} \emph {et~al.},\ }\bibfield  {title} {\bibinfo {title} {Error-corrected {H}adamard gate simulated at the circuit level},\ }\href {https://doi.org/10.22331/q-2024-07-02-1394} {\bibfield  {journal} {\bibinfo  {journal} {{Quantum}}\ }\textbf {\bibinfo {volume} {8}},\ \bibinfo {pages} {1394} (\bibinfo {year} {2024})}\BibitemShut {NoStop}%
\bibitem [{\citenamefont {Boixo}\ \emph {et~al.}(2018)\citenamefont {Boixo}, \citenamefont {Isakov}, \citenamefont {Smelyanskiy}, \citenamefont {Babbush}, \citenamefont {Ding}, \citenamefont {Jiang}, \citenamefont {Bremner}, \citenamefont {Martinis},\ and\ \citenamefont {Neven}}]{boixo2018characterizing}%
  \BibitemOpen
  \bibfield  {author} {\bibinfo {author} {\bibfnamefont {S.}~\bibnamefont {Boixo}}, \bibinfo {author} {\bibfnamefont {S.~V.}\ \bibnamefont {Isakov}}, \bibinfo {author} {\bibfnamefont {V.~N.}\ \bibnamefont {Smelyanskiy}}, \bibinfo {author} {\bibfnamefont {R.}~\bibnamefont {Babbush}}, \bibinfo {author} {\bibfnamefont {N.}~\bibnamefont {Ding}}, \bibinfo {author} {\bibfnamefont {Z.}~\bibnamefont {Jiang}}, \bibinfo {author} {\bibfnamefont {M.~J.}\ \bibnamefont {Bremner}}, \bibinfo {author} {\bibfnamefont {J.~M.}\ \bibnamefont {Martinis}},\ and\ \bibinfo {author} {\bibfnamefont {H.}~\bibnamefont {Neven}},\ }\bibfield  {title} {\bibinfo {title} {Characterizing quantum supremacy in near-term devices},\ }\href {https://doi.org/10.1038/s41567-018-0124-x} {\bibfield  {journal} {\bibinfo  {journal} {Nature Physics}\ }\textbf {\bibinfo {volume} {14}},\ \bibinfo {pages} {595} (\bibinfo {year} {2018})}\BibitemShut {NoStop}%
\end{thebibliography}%

\end{document}